\documentclass[twocolumn]{aastex631} %

\usepackage{multirow}
\usepackage{booktabs}
\usepackage{amsmath}
\usepackage{xcolor}
\usepackage{soul}

\DeclareRobustCommand{\add}[1]{\ifmmode{{\color{blue}#1}}\else{{\color{black} #1}}\fi}

\begin{document}

\title{The Stellar ``Snake''-V: a catalogue within 3\,kpc in the solar neighbourhood}

\author{Xiang-Ming Yang}
\affiliation{School of Mechanical Engineering, Hangzhou Dianzi University, Hangzhou 310018, China} 
\affiliation{Zhejiang Branch, National Astronomical Data Center, Hangzhou 310018, China}

\author{Ju-Yong Zhang} 
\affiliation{School of Mechanical Engineering, Hangzhou Dianzi University, Hangzhou 310018, China} 
\affiliation{Zhejiang Branch, National Astronomical Data Center, Hangzhou 310018, China}

\author[0000-0001-9289-0589]{Hai-Jun Tian$^*$}
\affiliation{School of Physics, Hangzhou Dianzi University, Hangzhou 310018, China} \email{hjtian@hdu.edu.cn}
\affiliation{Zhejiang Branch, National Astronomical Data Center, Hangzhou 310018, China}
\affiliation{Institute of Astronomy, Cambridge University, Cambridge CB3 0HA, UK} \email{ht525@cam.ac.uk}




\begin{abstract}
We present a Gaia DR3 source-level catalogue of \emph{Stellar Snake} complexes within 3\,kpc of the Sun. We define a Stellar Snake as a mutually coherent association of two or more stellar overdensities, characterised by consistent positions, kinematics, orbital invariants,
and ages, with broadly consistent chemical properties expected, rather than as a single gravitationally bound object. Moving beyond catalogue-driven searches seeded by known open clusters, our framework operates directly on individual Gaia sources to recover extended, low-density substructures and interconnecting stellar bridges. The multi-stage pipeline extracts statistically significant,
non-overlapping base nodes, infers homogeneous parameters with a
PointNet point-cloud regressor, and links them into large-scale
macro-structures using nine coordinates: position, tangential and
radial velocities, age \(\log t\), and \((E,L_Z)\). After FoF-topology consistency checking and boundary resolution,
the final catalogue contains 1,180 Stellar Snake candidates
with 4,196 final base nodes and 692,408 unique member-star entries,
of which \add{24.3\% are bridge stars}. Derived parameters are compared with external open-cluster
catalogues and spectroscopic benchmarks. To quantify structural coherence, we introduce a graph-relation
Snake Reliability Index (SRI) with recommended
Gold/Silver/Bronze grades. A separate spatial--kinematic partition diagnostic identifies
candidate subdivisions and provides separate scores for uniquely
selected cores. At the population level, the catalogue shows a descriptive
age--metallicity pattern in the jointly inferred parameters and a declining upper envelope of observed member-star entries
toward older inferred ages. \add{Young Snake nodes are not significantly closer to nearby
spiral-arm centrelines than matched young clusters, while those
selected near the projected Radcliffe Wave show a corresponding
vertical trend.}
\end{abstract}




\section{INTRODUCTION}
\label{sec:intro}

Stellar complexes are among the largest hierarchical structural units of star
formation in galactic disks. In the classical picture, they are spatially extended, kinematically coherent aggregates of OB associations and open clusters (OCs), formed within giant molecular cloud (GMC) complexes over characteristic timescales of $10^7$--$10^8$~yr \citep{efremov1995star, efremov1998hierarchical, elmegreen2009nature}. This hierarchical view implies that a broad census of nearby star formation
should not be limited to compact OCs, but should also recover loose
associations, extended substructures, and low-density bridges connecting
different levels of the star-formation hierarchy.

Before the \textit{Gaia} era, such a census was difficult to construct. Kinematic moving groups and OB associations provided important early evidence for dispersed stellar populations, but their interpretation was limited by astrometric precision, restricted distance coverage, and difficulty in distinguishing co-natal structures from dynamical resonances \citep{famaey2005local, antoja2008origin, de1999hipparcos}. As a result, the stellar-complex regime between compact clusters and large-scale Galactic structure remained only partially mapped.

The \textit{Gaia} mission has transformed this situation by enabling systematic searches for stellar groups across the Galactic disk. Large OC catalogues and blind clustering searches have added thousands of new clusters \citep{cantat2020painting, castro2022hunting, hunt2023improving}. In parallel, representative Gaia-based studies have expanded the open-cluster census in several complementary regimes, including early candidate searches, Galactic-disk cluster catalogues, nearby open clusters, and hidden stellar aggregates beyond the most complete nearby volume \citep{liu2019catalog, li2022lisc, qin2023hunting, he2023unveiling}. Open clusters have also been used as tracers of larger-scale Galactic structure \citep{hao2021evolution}, while cluster morphology provides an additional diagnostic of dynamical evolution and environmental effects \citep{hu2021decoding}.

These studies demonstrate the power of \textit{Gaia} for completing the OC census and for connecting open clusters to Galactic structure and dynamical evolution. However, most existing searches remain optimised for cluster-scale overdensities or catalogue-level cluster candidates. They are therefore less suited to recovering the full hierarchy of extended stellar complexes, in particular loose associations and low-density bridges between compact cluster nuclei. At larger spatial scales, \citet{kounkel2020untangling} extended density-based clustering to 5D phase space and identified extended co-moving populations known as Theia strings. However, the physical interpretation of these strings has been questioned: \citet{zucker2022disconnecting} showed that some candidates do not become tighter with improved astrometry, have velocity dispersions inconsistent with bound structures, and exhibit chemical homogeneity reproducible by random sampling. This concern highlights the need for more conservative searches that combine high-dimensional kinematic information with independent validation, rather than relying solely on present-day spatial and proper-motion coherence.

In response, recent work has increasingly adopted a cluster-group framework, in which confirmed OCs are linked using spatial, kinematic, dynamical, or orbital criteria \citep{liu2025formation, palma2025binary, swiggum2024most}. These studies provide a more physically interpretable route to identifying related cluster systems. Nevertheless, they inherit a structural limitation from their catalogue-driven design: candidate complexes whose constituent stars fall below current OC detection thresholds may remain invisible, and fixed linking criteria may sever extended low-density bridges between catalogued nuclei. A source-level search, beginning from individual Gaia stars rather than from known OCs, is therefore needed to probe the full multi-scale hierarchy of nearby stellar complexes. Nearby large-scale structures such as the Radcliffe Wave \citep{alves2020galactic} further motivate searches for stellar fossil counterparts of large-scale star-forming structures.

Within this context, the Stellar ``Snake''\footnote{https://astrotian.github.io/snake} has emerged as a representative example of an extended, kinematically coherent young cluster complex. Following its initial detection \citep{tian2020discovery}, \citet[Paper\,I,][]{wang2022stellar}
showed that the Snake extends to \(\sim500\) pc and contains 13 known OCs,
and identified four kinematic groups within the system.
\citet[Paper\,II,][]{yang2024stellar}
further resolved its internal partitioning and
identified a systematic head--tail asymmetry in its mass function,
consistent with feedback-mediated star formation. \citet[Paper\,III,][]{li2026stellar} extended the Snake framework to a younger, gas-rich system and showed that cloud density and early stellar feedback jointly regulate star formation. Most recently, \citet[Paper\,IV,][]{zhang2026emergence} used GALAH~DR4 spectroscopy to report early lithium depletion in the original Snake, demonstrating its value for stellar-evolution studies as well as clustered star formation.

Together, these studies show that extended stellar systems can be identified, dynamically connected, linked to their star-forming environments, and used to test stellar-evolution physics. They also demonstrate that the Snake concept, originally defined for
a single structure, can be generalised:
a Stellar Snake is an association of two or more distinct stellar
overdensities, characterised by consistent positions, kinematics,
orbital invariants, and ages, with broadly consistent chemical
properties expected. The term overlaps with stellar complexes,
cluster groups and families, primordial cluster groups, and, more
loosely, stellar strings. It does not require the entire system to
be gravitationally bound, elongated, strictly coeval, or demonstrated
to originate from a unique birth site. Our source-level pipeline
provides one operational implementation of this definition. What remains missing is the population-level analogue: a systematic catalogue of such stellar snake complexes across the broader solar neighbourhood. This work, Paper\,V of the series, is designed to fill that gap.

To construct such a catalogue, we adopt a source-level-to-complex strategy. The search volume is extended to several kiloparsecs from the Sun, comparable to the scale of nearby Galactic-disk structures. The search begins from individual Gaia sources rather than from a pre-existing OC catalogue, helping low-density substructures and bridges remain detectable. Candidate macro-structures are then identified in a feature space that combines present-day positions and kinematics with stellar ages and approximately conserved orbital quantities, such as the orbital energy \(E\) and vertical angular momentum \(L_Z\). Finally, the resulting associations are checked
for explicit conflicts with the original FoF hierarchy.

In this work, we apply this framework to Gaia DR3 sources within 3.1~kpc of the Sun, treating this as a 100 pc boundary buffer around a nominal 3 kpc survey volume. We first construct a FoF hierarchical tree and extract statistically significant, non-overlapping stellar nodes through local significance filtering and hierarchical de-duplication. We then use a customised PointNet point-cloud regressor to infer homogeneous node parameters, including age, distance, extinction, and metallicity. These base nodes, with homogeneous inferred parameters, are linked into macro-structures in a nine-coordinate physical and kinematic feature space using HDBSCAN, and the resulting candidates are
checked for explicit conflicts against the original FoF hierarchy. To assess spatial--kinematic coherence, we assign each finalised
structure a Snake Reliability Index (SRI) and a Gold/Silver/Bronze
quality flag. A separate partition diagnostic identifies
candidate subdivisions for further examination. The final catalogue contains 1,180 Stellar Snake candidates comprising
692,408 member-star entries.

The paper is organised as follows. Section~\ref{sec:data} describes the Gaia DR3 data selection and preprocessing. Section~\ref{sec:methodology} presents the full source-level-to-complex pipeline, including FoF node extraction, PointNet parameter estimation, HDBSCAN macro-structure reconstruction,
FoF-topology consistency checking,
and SRI reliability scoring. Section~\ref{sec:results} presents the final catalogue, reliability statistics, parameter validation, global demographic trends, and spatial distribution of the identified Stellar Snake candidates. Section~\ref{sec:discussion} compares the catalogue with external cluster systems, examines the kinematic and chemical coherence of base nodes, and revisits representative Snake systems. Section~\ref{sec:summary} gives the main conclusions, limitations, and future prospects.

\section{Data}
\label{sec:data}

This study is based on the high-precision astrometric data from Gaia Data Release 3 \citep[DR3;][]{vallenari2023gaia}. To search for Stellar Snake structures and their associated substructures within the solar neighbourhood, we select sources with direct inverse-parallax distances satisfying \(d_\varpi = 1/\varpi \leq 3.1~{\rm kpc}\), equivalently \(\varpi \gtrsim 0.32~{\rm mas}\). Although the intended survey volume is 3~kpc, we extend the selection boundary by an additional $\sim 100$~pc to 3.1~kpc. This buffer reduces the risk that structures lying close to the nominal 3~kpc limit are artificially truncated by boundary effects during clustering and significance analysis.

Only sources with a parallax signal-to-noise ratio
\(\varpi / \sigma_\varpi > 10\) are retained.
The resulting parent sample contains 94,133,866 Gaia DR3 sources.
This cut introduces a distance-dependent selection, increasingly
excluding faint sources at larger distances.

For the calculation of Galactocentric phase-space coordinates, orbital energy, vertical angular momentum, and orbit integrations, we adopt a single self-consistent solar-frame convention throughout this work. The solar Galactocentric radius is set to $R_0=8.27$ kpc, following the Galactic-rotation analysis of \citet{schonrich2012galactic}. We set the Sun's height above the Galactic mid-plane to $Z_\odot=25$ pc, corresponding to the reference value $z_0=25\pm5$ pc adopted in the Milky Way review of \citet{bland2016galaxy} and based on the SDSS star-count analysis of \citet{juric2008milky}. We adopt the solar peculiar motion with respect to the local standard of rest from \citet{tian2015stellar}, $(U_\odot,V_\odot,W_\odot)=(9.58,10.52,7.01)~{\rm km~s^{-1}}$, together with a local circular speed $V_c=238~{\rm km~s^{-1}}$ from \citet{schonrich2012galactic}. This corresponds to a total solar azimuthal velocity of $V_{\phi,\odot}=V_c+V_\odot=248.52~{\rm km~s^{-1}}$. The same $(R_0,Z_\odot,V_c)$ convention is used in the coordinate transformation, in the scaling of the Galactic potential, and in the calculation of the orbital energy $E$ and the vertical angular momentum $L_Z$. Orbit integrations are performed in a MWPotential2014-like axisymmetric Galactic potential \citep{bovy2015galpy}, rescaled to the same $(R_0,V_c)$ convention. With this normalisation, $L_Z$ is in kpc\,km\,s$^{-1}$ and $E$ in km$^2$\,s$^{-2}$. Since \(E\) depends on the adopted Galactic potential and both
\(E\) and \(L_Z\) depend on the solar reference frame, we use them as
internally consistent dynamical coordinates rather than as absolute
quantities to be compared across different conventions.

\section{Method}
\label{sec:methodology}

\begin{figure*}[htpb]
    \centering
    \includegraphics[width=0.65\textwidth]{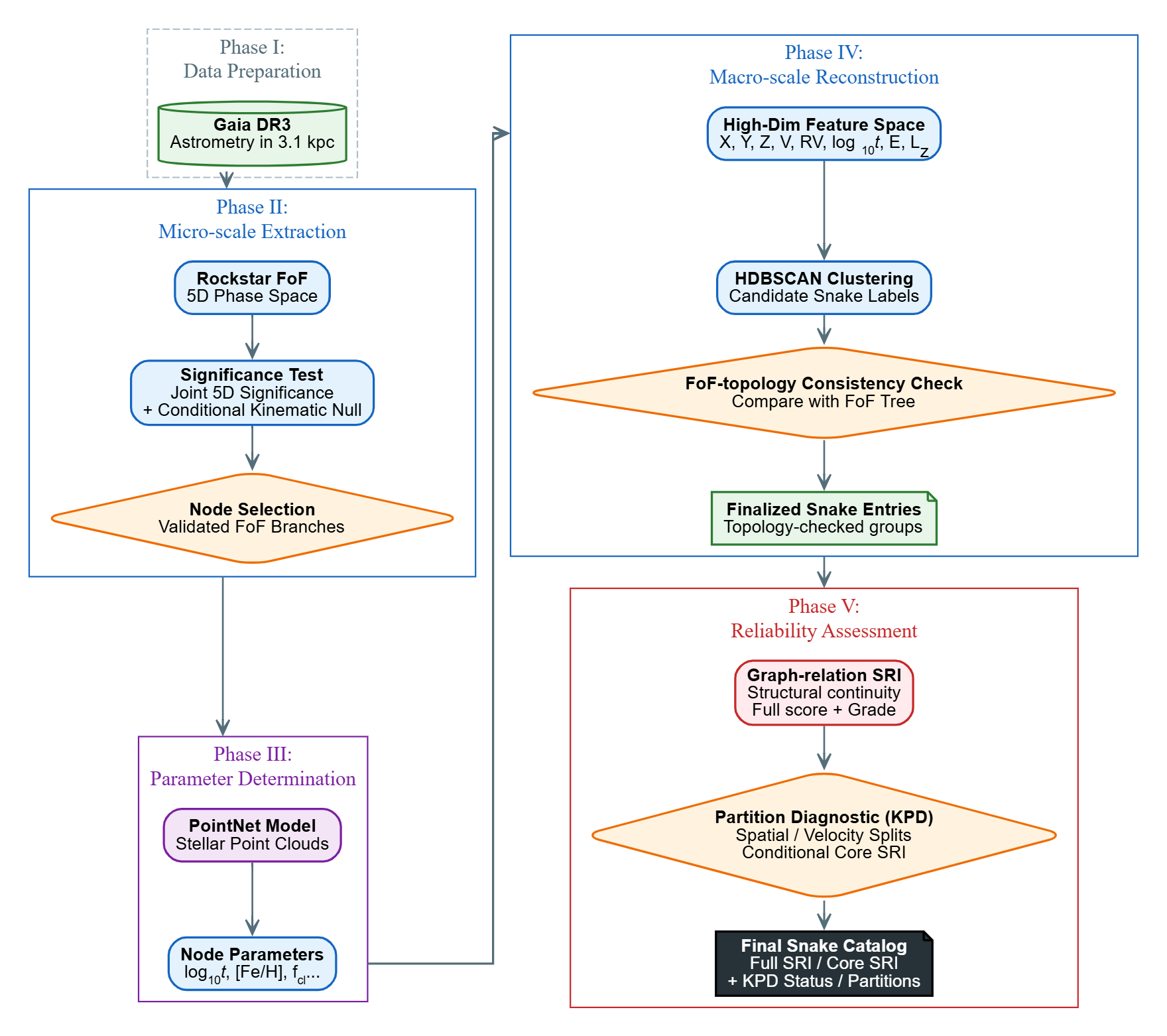}
    \caption{Schematic overview of the Stellar Snake identification pipeline.
    The workflow comprises five phases:
    (I) Gaia DR3 data preparation;
    (II) extraction of statistically significant micro-scale nodes via
    FoF hierarchical clustering;
    (III) physical parameter inference for each retained base node using
    a PointNet-based point-cloud regressor;
    (IV) macro-structure reconstruction in a
    nine-coordinate physical and kinematic feature space
    via HDBSCAN, followed by FoF-topology consistency checking
    against the original FoF tree;
    and (V) structural-coherence assessment through the SRI,
    with recommended Gold/Silver/Bronze grades and a separate
    kinematic-partition diagnostic (KPD).
    KPD identifies candidate spatial or velocity subdivisions;
    a core SRI is reported for flagged systems when a unique,
    scorable core can be selected.
    KPD and core scores do not alter catalogue membership or
    the grade assigned from the full SRI.}
    \label{fig:workflow}
\end{figure*}

To systematically identify and characterise extended Stellar ``Snake'' complexes in the solar neighbourhood, we develop a multi-phase data processing and machine-learning pipeline that links source-level \textit{Gaia} data to candidate, physically interpretable macro-structures. As shown in the workflow diagram (Figure~\ref{fig:workflow}), the methodology consists of five connected phases.

Throughout this work, we define a \emph{base node} as a statistically significant phase-space overdensity extracted from the FoF hierarchy. Such nodes are not required to be gravitationally bound; instead, they span the continuum from compact open clusters to dissolving or unbound associations. In the present catalogue, a \emph{Stellar Snake} is represented as
a graph-like assembly of two or more extracted base nodes, linked
through their positions, kinematics, ages, and orbital invariants. Chemical coherence remains part of the physical Snake concept, but in this catalogue it is evaluated as an independent consistency diagnostic rather than used as a coordinate in the fiducial linking step. Our physical interpretation concerns the relationships between nodes, not the internal boundedness of any individual node or member star.

The pipeline proceeds through five phases. \textbf{Phase I (Data Preparation; Section~\ref{sec:data})} defines the input \textit{Gaia} DR3 astrometric and photometric sample within the adopted search volume. \textbf{Phase II (Micro-scale Node Extraction; Sections~\ref{sec:fof}--\ref{sec:sel_node})} applies a Friends-of-Friends (FoF) hierarchical clustering algorithm, combined with local density-significance filtering, to identify statistically significant stellar nodes from the field background.

\textbf{Phase III (Node Parameter Estimation; Sections~\ref{sec:synthetic_data}--\ref{sec:pointnet})} employs a customised PointNet point-cloud regressor, trained on synthetic stellar systems, to estimate homogeneous physical parameters for each significance-filtered base node.

In \textbf{Phase IV (Macro-structure Reconstruction and
FoF-topology Consistency Checking;
Sections~\ref{sec:hdbscan}--\ref{sec:topological_validation})},
the parameterised nodes are projected into a
nine-coordinate physical and kinematic feature space
that includes positions, velocities, ages, and
derived dynamical quantities \((E, L_Z)\).
HDBSCAN is then applied to identify extended macro-structure candidates,
which are subsequently checked for explicit conflicts with the
original FoF hierarchy to resolve topological conflicts and assemble
the final catalogue.

Finally, \textbf{Phase V (Reliability Scoring and Quality Flags;
Section~\ref{sec:snake_reliability})} assigns each final structure
a graph-relation SRI and a Gold/Silver/Bronze quality flag.
A separate kinematic-partition diagnostic identifies
candidate spatial or velocity subdivisions and, where a unique
core can be selected, provides a separately reported core score.

The detailed mathematical framework and algorithmic implementation of each phase are described in the following subsections.

\subsection{Clustering Algorithm}
\label{sec:fof}

Within the adopted 3.1~kpc volume, we construct a hierarchical
FoF tree using the stellar phase-space adaptation of
\textsc{Rockstar}
\citep{behroozi2013rockstar,tian2017hide,tian2020discovery,wang2022stellar}.
The initial spatial grouping uses
\texttt{FOF\_LINKING\_LENGTH}=0.28 and
\texttt{AVG\_PARTICLE\_SPACING}=0.12207 in the adopted input scaling,
corresponding to a linking radius of approximately 48.8~pc.
This is a coarse pre-grouping scale, not a size criterion for a
final base node or Snake.

Each group is recursively refined using the native six-component
storage representation
\[
(X,Y,Z,RV_{\rm FoF},\mu_{\alpha *},\mu_\delta),
\qquad RV_{\rm FoF}\equiv0.
\]
The zero is a storage placeholder, not an estimate of a source's RV;
only five coordinates carry differential information.
Proper motions remain in the measured equatorial frame, without
conversion to tangential velocities or solar-reflex correction.
 With a single within-group proper-motion dispersion,
pairwise component distances are not generally invariant under
a change between equatorial and Galactic tangent-plane bases
over a wide field.

At each refinement, the position and motion subspaces are
normalised by their respective within-group dispersion scales,
with equal weights, \(w_{\rm pos}:w_{\rm pm}=1:1\).
\texttt{FOF\_FRACTION}=0.9 sets the quantile of within-group
nearest-neighbour distances used for the next linking scale.
Only subgroups with more than
\texttt{MIN\_HALO\_PARTICLES}=30 sources undergo further recursive
refinement; this is not a final-node richness cut.
We retain subgroup memberships, hierarchy levels, particle identifiers,
and stored coordinates at every recursive level, before the final
halo-output selection. The particle identifiers are mapped back
to Gaia source identifiers through the input catalogue. No gravitational-boundness criterion is imposed on this
retained stellar hierarchy. In particular,
\texttt{UNBOUND\_THRESHOLD}=0 removes the positive bound-fraction
requirement of the standard halo-output filter; it should not be
interpreted as disabling every internal unbinding calculation.
These are fixed operational settings, not uniquely optimal physical
scales for Galactic-disk complexes.
\add{Parallax-induced line-of-sight broadening can contribute to
larger measured three-dimensional node radii at greater distances;
within-group normalisation does not remove this effect.}

The complete FoF hierarchy contains 91,462 nodes. Since FoF can produce spurious linkages in low-density regions, we apply a statistical significance framework uniformly to these candidate nodes to suppress random associations and retain significant candidates for subsequent analysis.

The framework quantifies the phase-space density contrast between each candidate and its local environment. Following \citet{hunt2021improving}, which uses a Mann--Whitney U test to compare nearest-neighbour density distributions between candidate clusters and field stars, we adopt a similar philosophy for our hierarchical FoF nodes. Specifically, we use the 10th non-self nearest-neighbour distance, $D_{10}$, in the five-dimensional Mahalanobis space as the local-density statistic. For a candidate containing $N_C$ sources, we compare its $D_{10}$ distribution with an equal-size bootstrap sample of $N_C$ sources drawn from the complete local field pool, repeat the comparison 500 times, and convert the median one-sided Mann--Whitney $p$-value to a Gaussian-equivalent significance, $\sigma_{\rm obs}$. Because this significance retains a measurable dependence on richness, the $5\sigma$ threshold is used as an operational node-level criterion rather than as a catalogue-wide false-positive calibration.

To test whether the joint five-dimensional significance can be
explained by spatial concentration alone, we use a local conditional
kinematic null. Candidate positions are kept fixed. For each member,
a paired $(\mu_{\alpha*},\mu_\delta)$ is drawn uniformly from its
10 nearest eligible field stars, using a three-dimensional spatial
Mahalanobis metric estimated from the field. The nearest field star
to each candidate member is excluded from the donor pool, while the
evaluation field remains unchanged. Each null realisation is evaluated
using the same five-dimensional $D_{10}$ statistic and equal-size
Mann--Whitney procedure as the observed candidate.

A node is retained only when $\sigma_{\rm obs}>5$ and
$\sigma_{\rm obs}>\overline{\sigma_{\rm null}}+3s_{\rm null}$,
where $\overline{\sigma_{\rm null}}$ and $s_{\rm null}$ are the mean
and population standard deviation of the null significances.
The number of null realisations is increased adaptively until
omitting any single realisation leaves the acceptance decision unchanged.
Extreme-tail probabilities are evaluated without a fixed $p$-value
floor. Of the 91,462 nodes in the complete hierarchy, 61,806 pass
the first criterion and 61,638 satisfy both criteria, forming the
significance-filtered pool used for subsequent tree parsing.
In a semi-empirical control test, the conditional-null gate
rejects 1,168 of 1,172 cases retaining the observed spatial
configuration but with candidate-specific kinematic coherence
removed, while retaining all 612 corresponding positive controls.
This tests the intended discrimination rather than catalogue-wide
completeness or contamination.

\subsection{Extraction of Non-overlapping Base Nodes from the FoF Hierarchy}
\label{sec:sel_node}

Although 61,638 nodes pass the significance criteria, they still form
a deeply nested FoF hierarchy rather than a catalogue of mutually
independent structures. Within this connectivity graph, the same physical overdensity can appear multiple times as a sequence of ancestor, transitional, and descendant nodes. We therefore implement a graph-based tree-parsing and de-duplication procedure to extract a non-overlapping set of statistically significant stellar nodes for subsequent analysis.

The extraction procedure consists of three steps. First, in the \textit{Branch Potential Evaluation} step, we traverse the FoF hierarchy from the bottom upward and compute the maximum effective splitting depth for each node. This quantity measures whether a node contains descendant branches sufficiently deep to be treated as independent substructures. Second, in the \textit{Main-trunk Tracing and Bifurcation Identification} step, we start from the top-level root nodes and follow the longest contiguous branch downward. A node is treated as a bifurcation point when it splits into at least
two independent sub-branches with sufficient depth potential,
requiring each retained branch to persist across at least two
consecutive hierarchy levels, measured from its root or branch head
rather than a common absolute ROCKSTAR layer.
These sub-branches then serve as seeds for further tracing.
This two-level persistence requirement prevents terminal one-level
splits from being treated as independent branches; it is an operational
tree-parsing choice rather than a statistical significance threshold.

Finally, we apply a \textit{Redundant Parent Masking} step to avoid double-counting member stars. If a parent node bifurcates into descendant branches that are retained as independent nodes, the parent is treated as a transitional structure and excluded from the independent-node catalogue. By removing such retained transitional parents, this procedure yields a non-overlapping representation of the FoF hierarchy, ensuring that each retained node contributes a unique set of member stars to the subsequent analysis.

\begin{figure}[htpb]
\centering
\includegraphics[width=0.5\textwidth]{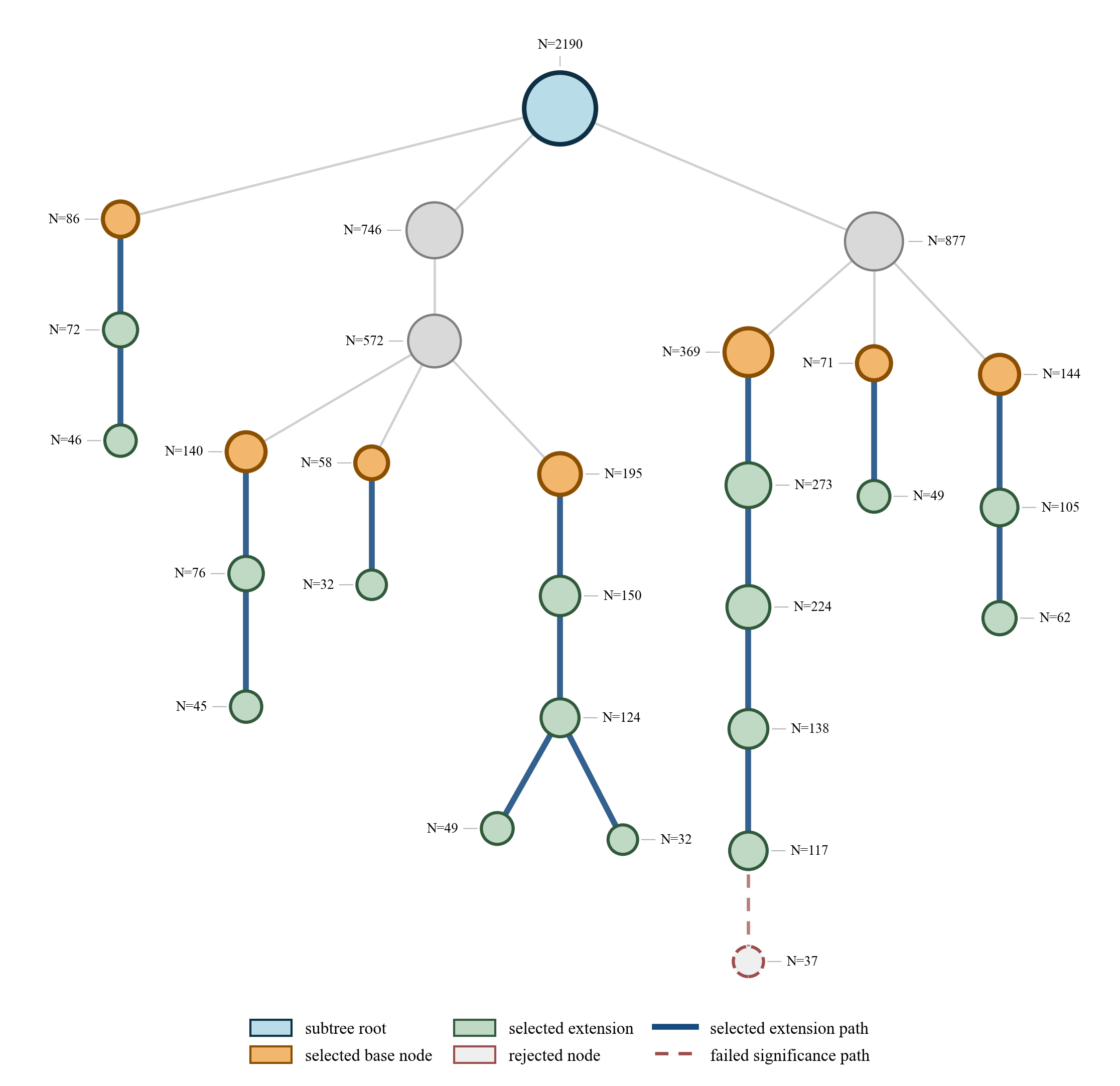}
\caption{Illustration of the hierarchical node-selection procedure for a representative FoF subtree. The light-blue node marks the subtree root. Orange nodes denote selected base nodes retained for subsequent analysis, and green nodes show their retained lower-level extensions. Grey nodes indicate other statistically significant nodes that are not retained as independent base nodes. Thick blue edges trace the retained extension branches attached to selected base nodes, while thin grey edges represent the remaining parent--child relations within the subtree. Nodes rejected by the statistical significance criteria are shown with dashed outlines and connected by dashed red edges. Symbol size scales with node member count \(N\), with values labelled.}
\label{fig:node_subtree}
\end{figure}

After this tree-parsing and de-duplication, the 61,638
significance-filtered hierarchical nodes are reduced to 9,966
non-overlapping base nodes. Figure~\ref{fig:node_subtree} illustrates this compression step for a representative FoF subtree. In this example, statistically significant nodes may still be rejected as redundant transitional parents when their descendant branches provide a more appropriate non-overlapping representation, while selected base nodes trace the branches that enter the final node representation. Branches that fail the statistical significance criteria are excluded.

These selected nodes should not be interpreted uniformly as classical open clusters or moving groups. The fine subdivision is intentional: we preserve statistically
supported substructures as separate units for subsequent association.
This design retains structural information for aggregation into
larger complexes rather than merging it at the node-extraction stage.  Rather, they form a mixed population of compact OCs, loose associations, and smaller substructures that satisfy the same statistical significance and de-duplication criteria. A key feature of this source-level construction is that the macro-structure search is not seeded by an external OC catalogue, but by stellar nodes identified directly from the Gaia data. These 9,966 base nodes therefore serve as the fundamental units for
subsequent parameter estimation and macro-structure reconstruction.

\subsection{Synthetic Training Set for the PointNet Regressor}
\label{sec:synthetic_data}

To train the PointNet regressor for node-level parameter estimation, we
generate 100,000 synthetic stellar populations, each representing a
cluster-like system with its own member stars. This synthetic set is designed
not to define the final macro-structures directly, but to expose the network
to a broad range of stellar-population properties and Gaia-like observational
degradation effects. The model can then infer homogeneous node-level
parameters---including age, distance, extinction, metallicity, photometric
broadening, and the photometric purity proxy \(f_{\rm cl}\)---for the
statistically selected FoF nodes.

The intrinsic stellar populations are generated from PARSEC isochrones
\citep{bressan2012parsec} combined with a Chabrier initial mass function
\citep{chabrier2014variations}. Following the stochastic synthetic-cluster
sampling strategy implemented in ASteCA \citep{perren2015asteca}, each
population is realised by drawing stars from the IMF and assigning unresolved
binaries probabilistically. Binary assignment follows the mass-dependent
prescription of \citet{duchene2013stellar}, in which the binary probability
depends on the primary-star mass. Interstellar extinction is parameterised by
\(A_V\) and applied to the Gaia passbands using colour- and
extinction-dependent coefficients,
\[
A_X = k_X\big((G_{\rm BP}-G_{\rm RP})_0, A_V\big)\,A_V,\quad
X\in\{G,G_{\rm BP},G_{\rm RP}\}.
\]
The coefficients adopt the Fitzpatrick extinction curve
\citep{fitzpatrick2019analysis} and follow the colour- and
\(A_V\)-dependent parametrisation of \citet{danielski2018empirical}, with a
fixed total-to-selective extinction ratio \(R_V=3.1\). The main input ranges
are summarised in Table~\ref{tab:synthetic_parameter_ranges}.

\begin{table}[!htbp]
\centering
\caption{Main input parameter ranges for the synthetic dataset.}
\label{tab:synthetic_parameter_ranges}
\begin{tabular}{lc}
\toprule
\textbf{Parameter} & \textbf{Sampling range} \\
\midrule
Input richness, \(\log_{10} N_*\) & \(\mathcal{U}(1.5,4.3)\) \\
Age, \(\log t\) & \([6.3,10.0]\) \\
Metallicity, \([{\rm Fe/H}]\) & \([-1.2,0.8]\) \\
Distance modulus, \(DM\) & \([3.0,12.5]\) \\
Visual extinction, \(A_V\) & \(\mathcal{U}(0.0,6.0)\) \\
Photometric purity proxy, \(f_{\rm cl}\) & \([0.05,1.0]\) \\
Photometric broadening, \(f_{\rm broad}\) & \(\mathcal{U}(1.0,20.0)\) \\
Richness-normalised radius, \(B_{50}\) & \([1.85,43.0]~{\rm pc}\) \\
Outer-envelope ratio, \(\eta_{98}=R_{98}/R_{50}\) & \([1.78,7.85]\) \\
Magnitude limit & \(G<18\,\rm mag\) \\
\bottomrule
\end{tabular}
\end{table}

Here \(f_{\rm cl}\) denotes a photometric purity proxy for the final
\(G<18\,\rm mag\) input point cloud. The nominal \(f_{\rm cl}\) range is \(0.05\)--1.0; the training
label is the actual synthetic-member fraction after field-star injection
and the final magnitude and colour--magnitude selections.

To avoid training only on idealised compact clusters, the spatial scale of
each synthetic system is drawn from a unified empirical projected-radius
prior. This prior is calibrated under the same \(G<18\,\rm mag\) selection
for both the Hunt--Reffert open-cluster catalogue \citep{hunt2024improving}
and our FoF base nodes. For both samples, \(R_{50}\) and \(R_{98}\) are
recomputed from the adopted centre, distance, and projected angular
separations. We then define the richness-normalised half-number radius
\(B_{50}=R_{50}/(N_*/100)^{1/3}\) and the outer-envelope ratio
\(\eta_{98}=R_{98}/R_{50}\). The main empirical calibration ranges are
\(B_{50}\in[1.85,43.0]~{\rm pc}\) and
\(\eta_{98}\in[1.78,7.85]\), with a small fraction of draws allowed to sample
the broader empirical tails.

For each synthetic system, the projected half-number radius is assigned as
\begin{equation}
R_{50}=B_{50}\left(\frac{N_*}{100}\right)^{1/3}.
\end{equation}
This prescription imposes a weak richness--size coupling, preventing the
training set from being dominated by unrealistic combinations such as very
rich but extremely compact systems or very sparse but excessively extended
systems. It is used only to define the synthetic training prior and is fixed
before applying the network to the observed nodes.

Because the PointNet input includes parallax broadening from line-of-sight
depth, we use the same characteristic scale to set a Gaussian depth
prescription. Specifically, the projected scale is mapped to a
HWHM-equivalent one-dimensional depth scale,
\begin{equation}
\sigma_{\rm los}=\frac{R_{50}}{\sqrt{2\ln 2}}.
\end{equation}
A small perturbation and a limited outer-envelope component are then added to
reproduce the observed spread in \(R_{98}/R_{50}\), while a geometric
positivity constraint prevents unphysical negative-distance tails. This
empirically calibrated radius prior spans compact open clusters as well as
loose and extended stellar aggregates, and exposes the network to the
non-Gaussian parallax signatures expected for spatially extended systems.

Observational degradation is tied directly to real Gaia DR3 data. The
baseline photometric uncertainties in \(G\), \(G_{\rm BP}\), and
\(G_{\rm RP}\) are inherited from observed sources, while parallax
uncertainties are sampled from an empirical magnitude-dependent error model.
This model is built by binning real Gaia sources in apparent \(G\) magnitude
and describing the parallax-error distribution in each bin with a robust
log-normal model, clipped to the empirical 1st--99th percentile envelope.
For the synthetic observations only, observed parallaxes are generated from
the geometric line-of-sight distances, the sampled parallax uncertainties,
and a constant approximation to the Gaia parallax zero point of
\(-0.017\) mas.
This constant mimics the mean zero-point offset of uncorrected
Gaia observations rather than the source-dependent correction of
\citet{lindegren2021gaia}. An assumed residual of
\(|\delta\varpi|\leq0.03\) mas corresponds to a distance-modulus
systematic of approximately \(-0.19\) to \(+0.21\) mag at 3.1~kpc,
with smaller effects at shorter distances.

Real Gaia colour--magnitude diagrams often show broader sequences than
expected from nominal measurement uncertainties alone. To capture this
mismatch, we introduce the empirical photometric broadening factor
\(f_{\rm broad}\) listed in Table~\ref{tab:synthetic_parameter_ranges}, which
rescales the baseline noise added to \(G\), \(G_{\rm BP}\), and
\(G_{\rm RP}\). The parallax uncertainties are not affected by this factor
and remain governed by the empirical magnitude-dependent parallax-error
model. We interpret \(f_{\rm broad}\) as a deliberately broad nuisance
parameter that absorbs residual broadening not captured by the nominal
uncertainty model; contributing effects may include differential reddening,
crowding, and stellar rotation, whereas unresolved multiplicity is modelled
separately. Held-out synthetic tests show estimator contraction
toward the lower \(f_{\rm broad}=1\) boundary.

Field contamination is modelled through a dynamic field-star injection
procedure. Instead of adding artificial noise points, we draw field stars from
the cleaned Gaia DR3 sample to construct a realistic background library. For
each synthetic system, contaminants are drawn from an adaptive local sky
region, with matching windows in parallax and proper-motion space. The
contamination fraction is defined as
\[
f_{\rm cont} = 1 - f_{\rm cl},
\]
so that the training set spans nearly uncontaminated systems as well as
heavily contaminated point clouds. The injected field stars are combined with
the synthetic member stars after applying the local measurement-error model.

The predicted photometric purity proxy \(f_{\rm cl}\) serves as an auxiliary
regression target and diagnostic quantity. Because its ground truth is defined
only for the synthetic training data, it is not used as a hard quality-control
criterion for the final catalogue.

\subsection{PointNet Architecture and Physics-informed Shortcut Connection}
\label{sec:pointnet}

The PointNet-based inference pipeline consists of two stages. First, two parallel PointNet encoders convert the unordered set of stars in each node into fixed-length global feature vectors through permutation-invariant operations. A parallax-aware encoder retains the per-star parallax information and feeds the physical-parameter heads, whereas a parallax-isolated colour--magnitude encoder feeds the \(f_{\rm cl}\) head. Second, a multi-task regression head combines these global representations with two node-level summary statistics---the median standardised \(\log_{10}\varpi_{\rm s}\) and the logarithmic richness---to predict the six node-level parameters, with a physics-informed shortcut supplying direct parallax information to the distance-modulus prediction.

\subsubsection{PointNet Encoder: Permutation-invariant Point Cloud Processing}

Traditional machine-learning approaches for extracting physical parameters from colour--magnitude diagrams (CMDs) often rely on Convolutional Neural Networks (CNNs), which require binning discrete stars into 2D pixelated histograms. This binning can introduce quantisation effects and complicates the incorporation of star-by-star astrometric measurements. To mitigate these limitations, we adopt and customise the PointNet architecture \citep{qi2017pointnet}. In our framework, each extracted stellar node is treated as an unordered set of stars, or a point cloud.

Each star is represented by a seven-dimensional effective feature vector combining Gaia photometry, parallax, colour, and derived colour--magnitude and parallax information:
\[
(G,\,G_{\rm BP},\,G_{\rm RP},\,\varpi,\,G_{\rm BP}-G_{\rm RP},\,
\log_{10}\varpi_{\rm s},\,D_{\rm CMD}),
\]
where \(\varpi_{\rm s}\equiv\max(\varpi,\,0.001\,{\rm mas})\). We denote by \(N_{\rm vis}\) the number of input stars passing the
\(G<18\) and colour--magnitude preprocessing selections.
Nodes with fewer than ten usable input stars are not assigned
PointNet parameter estimates. The quantity \(D_{\rm CMD}\) is a local sparsity measure in the colour--magnitude plane: within each node, the \((G,\,G_{\rm BP}-G_{\rm RP})\) plane is rescaled to \([0,1]\), and \(D_{\rm CMD}=\ln(\bar{d}+10^{-9})\), where \(\bar{d}\) is the mean of the \(k_{\rm eff}\) distances returned by the neighbour query, including the zero self-distance. We adopt
\begin{equation}
k_{\rm eff}=\max[3,\,\min(15,\,\lfloor 0.1\,N_{\rm vis}\rfloor)].
\end{equation}
Parallax is excluded from this neighbour search to avoid directly coupling \(D_{\rm CMD}\) to distance.

For fixed-size mini-batch training, each realisation randomly samples
512 stars from the node, with replacement for nodes having
\(N_{\rm vis}<512\).
For catalogue inference with \(N_{\rm vis}<512\), each usable
member is retained once and the remaining slots are filled by sampling
with replacement. Thus, 512 is an input-size convention, not a
minimum-member requirement. The per-star feature vector is augmented with three constructed PointNet coordinate channels \((G_{\rm BP}-G_{\rm RP},\,G,\,0)\) before entering the encoders. Because random subsampling does not by itself retain global node information, we append two node-level summary statistics after global pooling: the median standardised \(\log_{10}\varpi_{\rm s}\), computed before random subsampling, and the logarithmic richness \(\log_{10}(N_{\rm vis}+1)\).

The two encoders use different masked views of the same seven-dimensional feature set. In the parallax-aware branch, all seven per-star descriptors are retained. In the CMD branch, the per-star parallax and \(\log_{10}\varpi_{\rm s}\) channels are masked, leaving only \(G\), \(G_{\rm BP}\), \(G_{\rm RP}\), \(G_{\rm BP}-G_{\rm RP}\), and \(D_{\rm CMD}\). Thus the physical-parameter heads retain explicit parallax information,
whereas the \(f_{\rm cl}\) head is constrained to rely on CMD morphology
and richness rather than on explicit parallax information.
Apparent magnitudes and richness remain available, so this masking
does not imply statistical independence from distance or observational
quality.

Each encoder follows the standard PointNet design. Shared MLPs are applied independently to each star to map the input point into a high-dimensional latent feature space. A symmetric global max-pooling operation then aggregates the per-star feature matrix into a single global feature vector, ensuring permutation invariance. As in the original PointNet architecture, learned input and feature transformations are applied before global aggregation to align the input coordinate channels and intermediate feature space.

\subsubsection{Multi-task Regression with a Physics-informed Shortcut}

After global feature extraction, we restructure the original PointNet classification objective into a multi-task regressor. The network simultaneously predicts six node-level quantities: metallicity
\([{\rm Fe/H}]\), logarithmic age \(\log t\equiv\log_{10}(t/{\rm yr})\),
photometric broadening factor \(f_{\rm broad}\), visual extinction \(A_V\),
photometric purity proxy \(f_{\rm cl}\), and distance modulus \(DM\). Here \(f_{\rm broad}\) is the empirical scaling of the synthetic photometric
broadening introduced in Section~\ref{sec:synthetic_data}; \(f_{\rm cl}\) is
reported as a photometric purity proxy rather than used as a hard
quality-control criterion, because its ground truth cannot be independently
verified for every real FoF node.

The six outputs are produced by two isolated branches. The parallax-aware
encoder feeds a shared physical trunk that, together with the median
standardised \(\log_{10}\varpi_{\rm s}\) statistic, predicts
\([{\rm Fe/H}]\), \(\log t\), \(f_{\rm broad}\), \(A_V\), and a
distance-modulus residual. The parallax-isolated CMD encoder feeds the
\(f_{\rm cl}\) head, so that \(f_{\rm cl}\) is inferred from CMD morphology
and richness rather than from explicit parallax information. This routing makes \(f_{\rm cl}\) a photometric purity proxy rather
than a measure of parallax-space compactness.
The logarithmic richness \(\log_{10}(N_{\rm vis}+1)\) is supplied only to the
\(\log t\) and \(f_{\rm cl}\) heads, where source counts help constrain
main-sequence sampling and the apparent cleanliness of the CMD sample, and is
withheld from the metallicity, \(f_{\rm broad}\), \(A_V\), and distance heads
to prevent richness from acting as an unintended shortcut for quantities that
should be constrained by CMD morphology, photometric broadening, and parallax
information. This branch isolation and selective routing are intended to
preserve distance information for distance-dependent parameters while
preventing the \(f_{\rm cl}\) head from relying directly on parallax.

Because distance modulus inferred only from the learned point-cloud representation can be affected by the age--distance--extinction degeneracy, we introduce a physics-informed shortcut that supplies direct parallax information to the \(DM\) prediction. Specifically, the median standardised \(\log_{10}\varpi_{\rm s}\) is passed through a learnable linear layer and added to the physical-trunk \(DM\) residual in the normalised target space, before the prediction is transformed back to physical units. This additive formulation allows the physical trunk to learn residual corrections to the parallax-informed distance estimate and stabilises the joint prediction of distance-dependent parameters. The reported \(DM\) is therefore a parallax-anchored auxiliary
output, not an independent CMD-only distance estimate, and is not
used in HDBSCAN linking or the fiducial SRI.

For catalogue inference, each eligible base node is evaluated
100 times with dropout active, fixed Batch-Normalization statistics,
and member resampling to 512 inputs. The reported parameters and
\(e_*\) values are the means and standard deviations of these
predictions. The latter measure internal Monte-Carlo repeatability,
not externally calibrated total uncertainties or formal
\(1\sigma\) confidence intervals.

\subsection{Macro-structure Identification via Reliability-weighted HDBSCAN}
\label{sec:hdbscan}

The FoF procedure identifies statistically significant nodes in the 5D
astrometric space \((X,Y,Z,\mu_{\alpha*},\mu_\delta)\), which excludes
radial velocity. Because per-star radial velocities are available only for a subset of
\textit{Gaia} sources and source-level links across extended structures are
sensitive to projection effects, we assemble macro-structures at the node
level, using the 9,966 non-overlapping base nodes as the input pool. For
each node we form the feature vector
\[
[X,\,Y,\,Z,\,V_{\alpha*},\,V_\delta,\,RV,\,\log t,\,E,\,L_Z],
\]
where \(V_{\alpha*}\) and \(V_\delta\) are node-median tangential
velocities, \(RV\) is the node-level radial velocity, \(\log t\) is the
PointNet age, and \(E\) and \(L_Z\) are the orbital energy and vertical
angular momentum, computed in the solar-frame and potential convention of
Section~\ref{sec:data}. The nine coordinates include two derived
dynamical quantities, \(E\) and \(L_Z\), which re-weight existing
phase-space information. \add{Their inclusion as auxiliary coordinates
is intended to help reconnect finely resolved FoF fragments that may
remain close in integrals-of-motion space, following the suggestion
of \citet[Chapter~2]{tian2017hide}.} The spatial and tangential-velocity axes are
available for every node and form the backbone of the search. The remaining axes are heterogeneous: \(RV\) requires
line-of-sight measurements, \(E\) and \(L_Z\) inherit the \(RV\) sampling, and
\(\log t\) depends on a node's richness and colour--magnitude coverage. We therefore enter these auxiliary axes through a reliability-weighted
precomputed dissimilarity
rather than an equal-weight Euclidean distance.

Each axis is first transformed to a dimensionless median/IQR-scaled
coordinate using a robust scaler, and no dimensional reduction is applied.
Writing \(s\) for the scaled features
and \(\Delta_{ij,k}=s_{i,k}-s_{j,k}\), we define
the symmetric squared dissimilarity
\begin{equation}
\begin{aligned}
D_{ij}^{2}
&=
\frac{
\displaystyle
\sum_{c\in{\rm core}}\Delta_{ij,c}^{2}
+
\sum_{d\in{\rm aux}}w_d
\sum_{f\in d}\rho(r_{i,f},r_{j,f})\Delta_{ij,f}^{2}
}{
\displaystyle
5+\sum_{d\in{\rm aux}}w_d
\sum_{f\in d}\rho(r_{i,f},r_{j,f})
},
\end{aligned}
\label{eq:weighted_distance}
\end{equation}
with \({\rm core}=\{X,Y,Z,V_{\alpha*},V_\delta\}\) at unit weight and
\({\rm aux}=\{\log t,\,RV,\,(E,L_Z)\}\). The two modulating factors are
defined by fixed, data-determined rules: one measures the global
informativeness of each auxiliary block, and the other measures the
pairwise reliability with which that block is available for the two nodes.

The global informativeness weight uses the Hopkins clustering-tendency
statistic \(H_d\) \citep{hopkins1954new,banerjee2004validating}, evaluated
on the scaled node values of auxiliary block \(d\) using repeated
subsampling. We map it to
\begin{equation}
w_d=\mathrm{clip}\!\left[\,2\,(H_d-0.5),\,0,\,1\,\right],
\end{equation}
so that a block close to a random distribution
\((H_d\simeq0.5)\) receives \(w_d\simeq0\), whereas a strongly structured
block approaches \(w_d=1\). Once this functional form is specified, the
weights are computed from the node table and are not tuned separately for
individual regions or candidate structures.

The pairwise reliability factor is the harmonic mean
\begin{equation}
\rho(r_{i,f},r_{j,f}) =
\frac{2r_{i,f}r_{j,f}}{r_{i,f}+r_{j,f}},
\end{equation}
with \(\rho=0\) when \(r_{i,f}+r_{j,f}=0\).
Here \(f\) indexes a coordinate within auxiliary block \(d\).
The age reliability is the normalised empirical rank of the number
of member sources with \(G<18\). For \(f\in\{RV,E,L_Z\}\), we use
\(r_{i,f}=\sigma_{f,0}^{2}/(\sigma_{f,0}^{2}+\sigma_{i,f}^{2})\),
where \(\sigma_{i,f}\) is the effective node-level \(RV\) uncertainty
or the corresponding \(E\) or \(L_Z\) uncertainty, and
\(\sigma_{f,0}\) is its median over valid positive values.
\(E\) and \(L_Z\) retain separate reliabilities but share the global
orbital-block weight. Unavailable coordinates or invalid uncertainties
receive zero reliability and contribute to neither sum.

We adopt the median Hopkins statistic from 25 repeated subsamples,
each using 25\% of the valid nodes, restricted to 50--2000 objects.
The resulting \((H_d,w_d)\) values are
\((0.6894,0.3788)\) for age,
\((0.9946,0.9892)\) for \(RV\), and
\((0.9970,0.9941)\) for \((E,L_Z)\).
The scaling medians and IQRs are listed in
Table~\ref{tab:hdbscan_scaling}.
This scaling gives equal physical separations greater weight in
\(Z\) than in \(X\) or \(Y\).

\begin{table}[!htbp]
\centering
\caption{Median and interquartile-range scales of the HDBSCAN
input coordinates.}
\label{tab:hdbscan_scaling}
\begin{tabular}{lcrr}
\toprule\toprule
Coordinate & Unit & Median & IQR \tabularnewline
\midrule
\(X\)          & pc                  & 182.216    & 1205.194 \tabularnewline
\(Y\)          & pc                  & -31.785    & 1389.524 \tabularnewline
\(Z\)          & pc                  & -25.902    & 231.091  \tabularnewline
\(V_{\alpha*}\)& km\,s\(^{-1}\)      & -8.170     & 25.103   \tabularnewline
\(V_\delta\)   & km\,s\(^{-1}\)      & -13.162    & 23.831   \tabularnewline
\(\log t\)     & dex                 & 8.6125     & 1.3887   \tabularnewline
\(RV\)         & km\,s\(^{-1}\)      & -6.993     & 28.920   \tabularnewline
\(E\)          & km\(^{2}\)\,s\(^{-2}\)
                                    & -45680.020 & 8578.794 \tabularnewline
\(L_Z\)        & kpc\,km\,s\(^{-1}\) & 1894.262   & 287.881  \tabularnewline
\bottomrule
\end{tabular}
\end{table}

The pair-dependent dissimilarity is symmetric but need not satisfy
the triangle inequality. HDBSCAN density and membership quantities
are not interpreted as physical posterior probabilities. Unlike mean or median
imputation, this treatment assigns unmeasured nodes no common filler value
that could make them appear spuriously similar. Because only the five core axes must be finite, all 9,966 base nodes
enter the clustering.

We apply HDBSCAN \citep{mcinnes2017hdbscan} to the
precomputed dissimilarity matrix \([D_{ij}]\),
with \texttt{min\_cluster\_size}=2,
\texttt{min\_samples}=2, and leaf selection. The
\texttt{min\_cluster\_size}=2 choice follows the operational definition that
the smallest possible Stellar Snake candidate contains two associated base
nodes. The choice of \texttt{min\_samples}=2 with leaf extraction
favours finely resolved candidate groupings, preserving smaller
associations for subsequent assessment and higher-level aggregation.
A larger physical association may therefore remain distributed
across several catalogue entries. The physical reliability
of the resulting structures is assessed subsequently through FoF-topology
consistency checking and the SRI framework.

We retain nodes with HDBSCAN membership probability
\(P\ge0.7\), reset lower-probability nodes to noise, and remove
candidate labels left with fewer than two nodes.
At the adopted setting, this removes 110 of 4,038 non-noise
nodes (2.72\%). All members of the 412 two-node candidates have
\(P=1\), so none is removed. The cut is a boundary-support diagnostic,
not an effective rejection test for two-node associations or a
physical association probability.

The retained labels are candidate groupings, not final memberships.
They are checked against the FoF hierarchy in
Section~\ref{sec:topological_validation}, and the assembled structures
are subsequently assessed by the SRI and a separate partition diagnostic.

A sensitivity comparison using the same input nodes and
\texttt{min\_cluster\_size}=2 with leaf extraction gives 1,180,
511, and 328 final structures for \texttt{min\_samples}=2, 5,
and 10, respectively, retaining 42.10\%, 19.79\%, and 12.94\%
of the 9,966 input nodes. Under the same SRI prescription
(Section~\ref{sec:snake_reliability}), their median full scores are
0.747, 0.737, and 0.789, with 16th--84th percentile ranges of
0.586--0.887, 0.549--0.905, and 0.586--0.923.
The corresponding Gold/Silver/Bronze counts are
727/365/88, 301/155/55, and 216/82/30.
Only 91 and 98 higher-\texttt{min\_samples} entries have exactly the
same node sets as entries in the adopted catalogue.
These three final-catalogue comparisons show sensitivity to grouping
resolution; changing membership and evidence-group composition
prevents interpreting higher scores alone as improved physical purity.

\subsection{FoF-topology Consistency Checking and Final Catalogue Assembly}
\label{sec:topological_validation}

Although HDBSCAN generates macro-structure candidates in the
nine-coordinate node-level feature space, its groupings can be affected
by over-merging or fragmentation. We therefore check them against
the original FoF hierarchy before assembling the final catalogue.
The hierarchy is used to identify explicit topological conflicts,
trace node ancestry, and resolve boundaries, not as an independent
physical classifier.

The central operation is \textit{root-node tracing}. For each non-noise
HDBSCAN grouping, we identify the shallowest FoF parent branches that
cover its nodes without enclosing nodes assigned to another non-noise
grouping. Base nodes labelled as noise after the probability cut are
automatically included when they lie within these selected branches.
This automatic assembly provides the reference node set for identifying
subsequent manual modifications. Each final catalogue entry is assigned
to exactly one of four mutually exclusive provenance classes:

\begin{enumerate}
    \item \textbf{Same-root HDBSCAN--FoF agreement:}
    the final node set coincides with the automatic assembly of one
    non-noise HDBSCAN grouping within a single FoF root, without manual
    modification.

    \item \textbf{Cross-root HDBSCAN grouping without FoF contradiction:}
    the final node set coincides with an automatically assembled HDBSCAN
    grouping spanning two or more independent FoF roots, without manual
    modification. This path records the absence of an explicit FoF
    contradiction, not positive support for the cross-root association.

    \item \textbf{Conflict-resolved grouping:}
    the candidate enters the conflict-review procedure and its final
    membership or grouping is changed through merging, splitting, node
    removal, or reassignment.

    \item \textbf{FoF-only/noise-recovery grouping:}
    a single FoF root contains at least two retained nodes whose
    post-probability-cut HDBSCAN labels are all noise. These nodes are
    retained as a separate multi-node candidate only when they are
    mutually consistent and cannot be assigned to an existing non-noise
    HDBSCAN grouping.
\end{enumerate}

A case is flagged for conflict review when two or more HDBSCAN groupings
share the same FoF root, or when a single HDBSCAN grouping combines
branches that are distinct in the FoF hierarchy. The review is performed
at the FoF-root level using the three-dimensional spatial distribution,
the tangential kinematics, reliable radial-velocity information where
available, the colour--magnitude and age distributions, the orbital
quantities, and the internal FoF branch topology. Candidate labels are
merged only when their spatial--kinematic distributions are continuous
and the population and orbital diagnostics show no clear incompatibility.
They are split when distinct FoF branches are accompanied by a clear
spatial, kinematic, or orbital discontinuity. Nodes that are isolated
from or incompatible with the remaining structure may be removed.
An HDBSCAN-noise node is reassigned only when it has a consistent
hierarchical and spatial--kinematic association with an existing
structure. After the root-level decision, the remaining members of the
affected labels are re-examined and removed or separated when necessary.

A catalogue entry that enters conflict review but retains its original
membership is not classified as conflict-resolved. It retains the
appropriate automatic provenance class and is recorded with
\texttt{conflict\_reviewed}=1 and
\texttt{manual\_modified}=0. Entries whose membership or grouping is
actually changed are assigned the conflict-resolved provenance and
\texttt{manual\_modified}=1.

In total, 44 conflict-flagged FoF roots were reviewed, yielding
76 modified final catalogue entries. A root-level review can affect
multiple final entries.

After the FoF-topology consistency check and boundary-resolution
procedure, the final catalogue contains 1,180 Stellar Snake entries.
The contribution of each provenance route is reported in
Section~\ref{subsec:catalog_overview}, and the provenance and review
fields are defined in Table~\ref{table:snake_complex}. The final
entries, rather than the raw HDBSCAN labels, are used in all subsequent
catalogue-level analyses. The FoF-topology check resolves the most evident conflicts.
The spatial--kinematic coherence and possible subdivisions
of the assembled structures are assessed separately in the
following section.

Bridge stars are retained sources of a FoF part contributing
at least two retained base nodes to the same Snake, but are not assigned
to any retained base node. They inherit their FoF part's Snake assignment,
with valid \texttt{Snake} and \texttt{id\_part} identifiers and a blank
\texttt{id\_node}. This definition applies at the FoF-part level;
a shared root or HDBSCAN grouping alone is insufficient.

\subsection{Snake Reliability Index and Quality Flags}
\label{sec:snake_reliability}

The Snake Reliability Index (SRI) is a continuous post-assembly
measure of spatial--kinematic compatibility, not a probability of
physical association. Age, metallicity, orbital energy, and angular
momentum do not enter the score. Locally compatible base-node fragments
may first contract into graph entities through spatial-contact and
tangential-velocity criteria, with an RV-conflict veto where measurements
permit. Contraction changes only the scoring representation; if it
would leave one entity, the original base nodes are retained.

The five components are spatial and velocity member-cloud
continuity, \(S_{\rm mem}\) and \(V_{\rm mem}\), spatial and
velocity centre-discontinuity scores, \(S_{\rm cen}\) and
\(V_{\rm cen}\), and the global spatial--velocity relation score,
\(C_{\rm cross}\) (Appendix~\ref{app:sri}). Member-velocity comparisons
use a common tangent plane and model-based perspective transport only
when both endpoint entities have sufficient RV support; otherwise both
use the RV-free representation. The three-dimensional velocity centres
and their uncertainties are estimated separately from each entity's own
source measurements, without filling missing RVs. These representative
centres and the RV-free comparisons retain the geometric limitations
specified in the appendix.

Let \(N_{\rm entity}\) be the number of scoring entities and
\(M_{\rm RV}\) the number with finite three-dimensional centres supported
by at least three distinct sources with valid RVs and positive, finite
measurement uncertainties. We report evidence group A for
\(M_{\rm RV}<3\), B for \(M_{\rm RV}=3\), and C for
\(M_{\rm RV}\ge4\). Group A uses member scores alone. In Groups B and C,
the spatial and velocity pillars are the geometric means of the
corresponding member and centre scores. The spatial centre term uses
all spatially valid entities; the velocity centre term and Cross use
only RV-qualified entities. The pillars are combined by their harmonic
mean \(R_0\). Group C additionally uses a regularised leave-one-out
affine velocity prediction, giving \({\rm SRI}=R_0C_{\rm cross}\);
otherwise \({\rm SRI}=R_0\). The catalogue contains 552, 273, and 355
systems in evidence groups A, B, and C, respectively.

Gold, Silver, and Bronze are empirical recommended grades:
\(0.70\le{\rm SRI}\le1\), \(0.50\le{\rm SRI}<0.70\), and
\(0\le{\rm SRI}<0.50\), respectively. The continuous score is primary;
users may loosen or tighten these boundaries for their scientific
objectives while accounting for evidence group and measurement support.
The grades are not calibrated to a common false-positive rate
across evidence groups. The recommendations are informed by representative
systems and checked against the Palma pair benchmark
(Section~\ref{subsubsec:palma_oc_systems}); the optical-control
false-positive fractions are 1.72\% at 0.70 and 5.41\% at 0.50.
These benchmark rates are not catalogue-wide contamination rates or
evidence of equal purity across A/B/C.

A separate kinematic-partition diagnostic (KPD) identifies prominent
subdivisions through recursive longest-edge comparisons with a relative
threshold of 3 and cloud-width protection for spatial and RV-free
member-cloud comparisons. It does not use an SRI-gain requirement,
change the full SRI, or remove catalogue members. Systems with
\(N_{\rm entity}<3\) are unassessed, not unflagged. KPD is a separate
output from the same source data, not statistically independent evidence.

For a flagged system, the core is the uniquely selected partition
component with the most entities, with node-member count breaking ties.
A core containing at least two entities is rescored from its retained
sources while preserving that entity partition. Ambiguous assignments
or unresolved ties yield no core score. Core scores may use a different
evidence group and do not replace the full-score grade. For a restricted
sample of assessed systems without a detected subdivision, users can
require Gold and \({\rm KPD}=0\), while retaining KPD-unassessed Gold
systems as a separate population. Node-level FoF significance and
PointNet \(f_{\rm cl}\) provide additional diagnostics; neither enters
the SRI grade.

\begin{figure*}[t]
    \centering
    \includegraphics[width=0.92\textwidth]{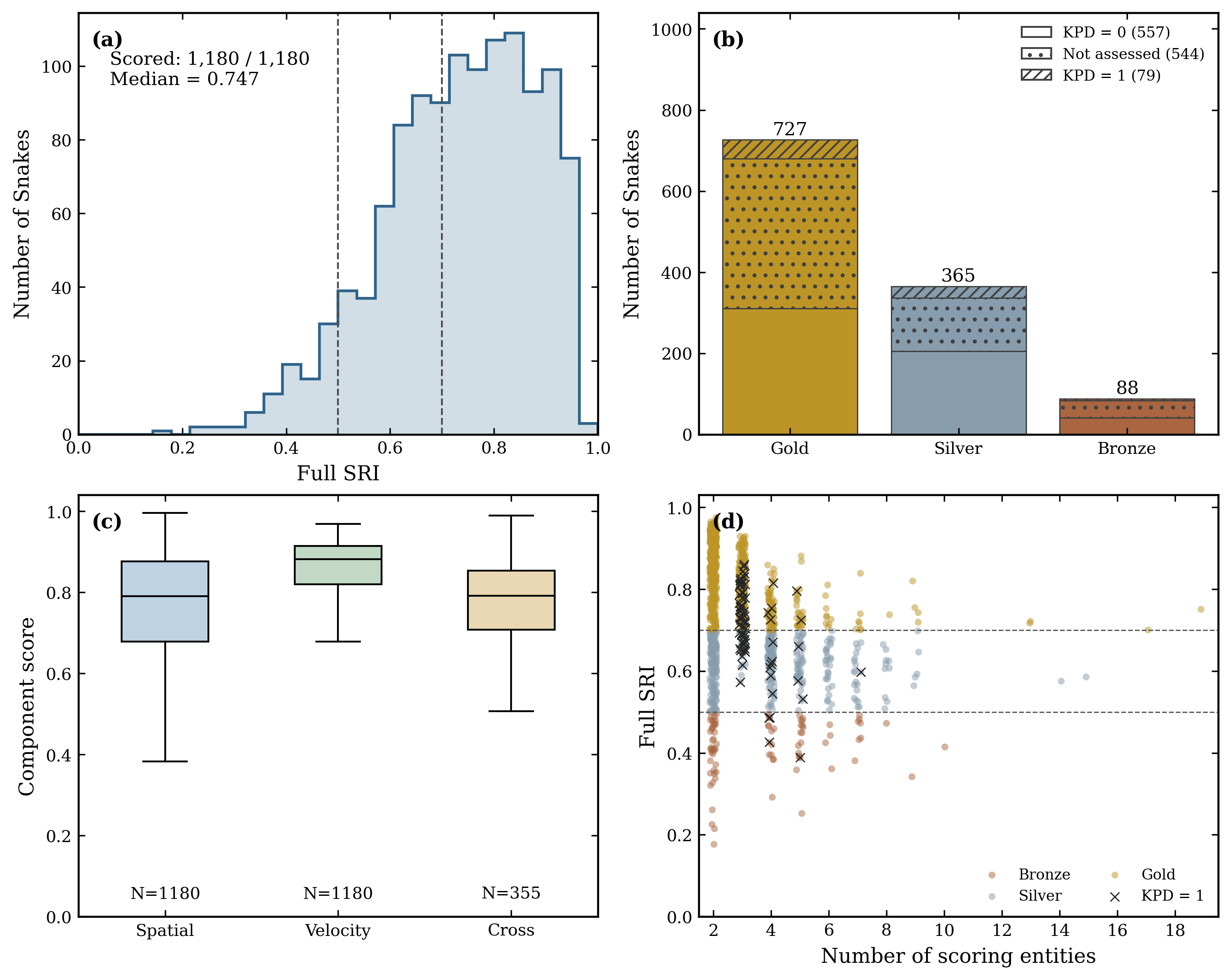}
    \caption{Full-system SRI and its diagnostics for all 1,180 Snakes.
    (a) Continuous SRI distribution; dashed lines mark the recommended
    boundaries at 0.50 and 0.70. (b) The 727 Gold, 365 Silver, and
    88 Bronze systems, separated into assessed KPD=0, assessed KPD=1,
    and unassessed populations. KPD is assessed for 636 systems,
    including 79 flagged systems; 544 are unassessed.
    The flag fraction is 12.42\% of assessed systems or 6.69\% overall.
    (c) Spatial and velocity pillar distributions for all 1,180 systems,
    and Cross for the 355 Group-C systems where it is active.
    (d) Full SRI versus scoring-entity count, with crosses marking KPD=1.
    Grades use the continuous full score;
    KPD and core scores are separately reported diagnostics.}
    \label{fig:sri_reliability_summary_graph}
\end{figure*}

All 1,180 systems have finite full scores: 727 Gold, 365 Silver,
and 88 Bronze, with median 0.747 and a 16th--84th percentile range of
0.586--0.887 (Figure~\ref{fig:sri_reliability_summary_graph}).
\add{The Gold fractions within evidence groups A, B, and C are
376/552 (68.12\%), 247/273 (90.48\%), and 104/355 (29.30\%),
respectively.}
KPD is assessed for 636 systems: 79 are flagged and 557 are unflagged;
544 are unassessed. The flag fractions are therefore 79/636
(12.42\%) among assessed systems and 79/1,180 (6.69\%) overall.
All 79 flagged systems have a computable core score in this catalogue.

\section{Results}
\label{sec:results}

In this section, we present the properties of the final Stellar Snake catalogue. We first summarise the overall content and sky coverage of the catalogue (Section~\ref{subsec:catalog_overview}), then assess the PointNet-derived node parameters against synthetic
and external benchmarks (Section~\ref{subsec:model_validation}). We next examine the global demographic trends of the Snake population (Section~\ref{subsec:global_properties}) and investigate their spatial association with nearby spiral-arm loci and the Radcliffe Wave (Section~\ref{subsec:spatial_distribution_spiral_arm}).

\subsection{Final Catalogue Overview and All-sky Distribution}
\label{subsec:catalog_overview}

After the HDBSCAN candidate-generation step,
FoF-topology consistency checking, and final boundary resolution
described in Section~\ref{sec:topological_validation}, we obtain a
finalised catalogue of 1,180 Stellar Snake candidates.
The catalogue contains 692,408 unique Gaia sources, including
168,232 bridge stars (24.30\%). Of the 9,966 input base nodes,
4,196 (42.10\%) enter the final Snake catalogue, while
5,770 (57.90\%) do not.

\paragraph{Catalogue provenance.}
The four mutually exclusive assembly routes defined in
Section~\ref{sec:topological_validation} comprise
21 same-root agreement entries (1.78\%),
1,025 cross-root entries without an explicit FoF contradiction (86.86\%),
76 conflict-resolved entries (6.44\%), and
58 FoF-only/noise-recovery entries (4.92\%).
Of the 113 final entries associated with conflict review,
76 were modified and 37 retained their automatic assembly,
as recorded by the catalogue review flags.

\paragraph{FoF-only/noise-recovery subset.}
Of the 61 initial FoF-only candidates, 58 remain after final boundary
and coherence review, while three are rejected.
Each retained entry contains at least two nodes from the same FoF root
whose post-probability-cut HDBSCAN labels are noise.
This subset is therefore reported separately from candidates supported
by a non-noise HDBSCAN grouping.

\add{In a comparison with the unnormalised calculation,
449 of the 1,180 adopted Snakes (38.05\%) lack a counterpart
with identical node and member-source sets.
The adopted and RUWE \(<1.4\) comparison catalogues share
44.1\% and 48.5\% of their member sources, respectively;
the adjusted Rand index on the shared sources is 0.920.
Table~\ref{table:snake_complex} provides each Snake's best-match
member overlap with the RUWE-selected catalogue.}

Each entry reports the continuous full SRI, its component scores,
the A/B/C evidence group, scoring-entity and RV-qualified-entity counts,
and KPD assessment status (Table~\ref{table:snake_complex}).
The recommended grades do not replace these diagnostics or prescribe
one threshold for all uses. In the Palma optical control, the default
0.70 and 0.50 thresholds select 1.72\% and 5.41\% of systems,
respectively, not calibrated fractions of false associations in this
catalogue (Section~\ref{subsubsec:palma_oc_systems}).
Core scores are available only for uniquely selected KPD cores and
must not be substituted for full scores without specifying the
changed sample and evidence group. The source-level observations
required for reproducible SRI calculation are retained in
Table~\ref{table:snake_member}.

\add{Snake~1078 and Snake~1085 contain subdivisions of the
globular clusters NGC~6121 and NGC~6397, respectively.
Their node-level PointNet estimates are retained for reference
and should be used with caution given the training-domain
limitations (Table~\ref{tab:synthetic_parameter_ranges}).}

\begin{figure*}[htpb]
    \centering
    \includegraphics[width=\textwidth]{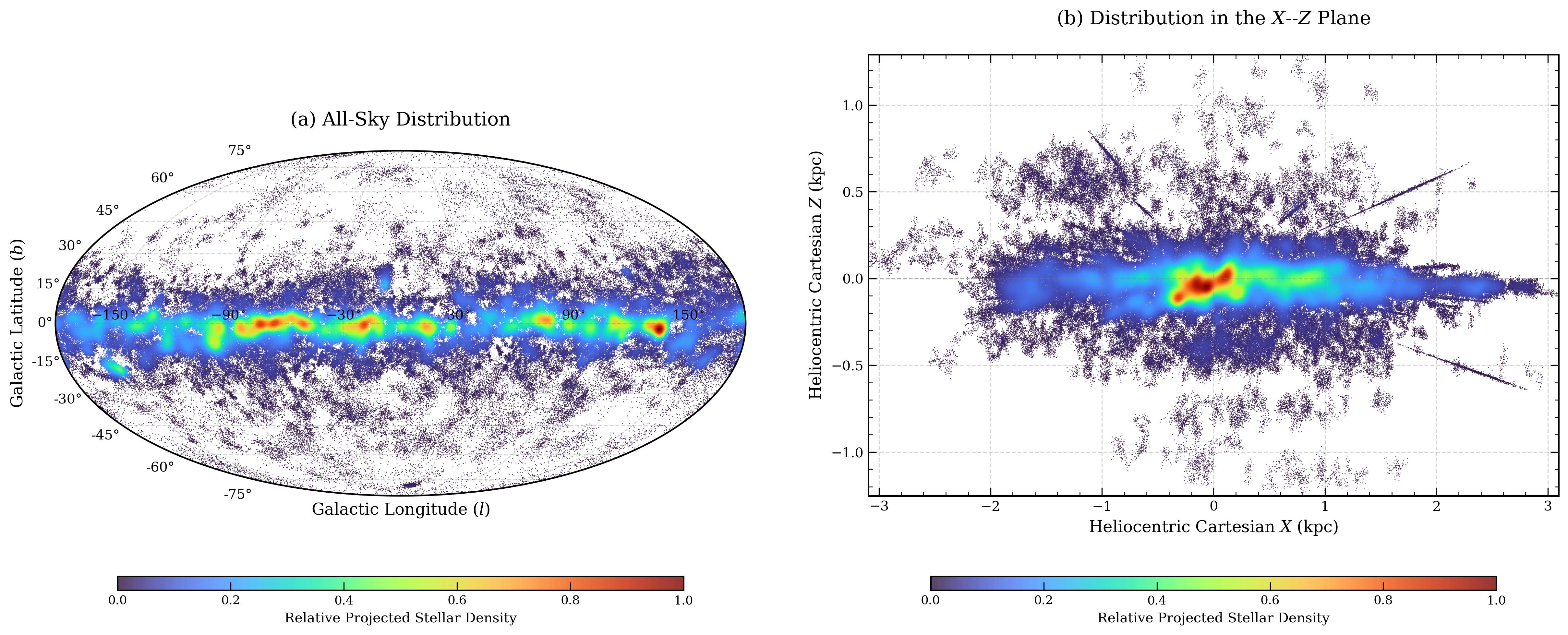}
    \caption{Source-level spatial distribution of the final Stellar Snake catalogue.
    Panel (a) shows the 692,408 unique Gaia member sources in Galactic coordinates
    \((l,b)\), while panel (b) shows the same sources in the heliocentric
    Cartesian \(X\)--\(Z\) plane. The sources belong to all 1,180 final catalogue
    structures. Colours indicate the relative projected member-source density and
    are normalised independently in the two projections, with warmer colours
    marking higher projected density. These projections show the sky coverage and
    the in-plane and vertical extent of the catalogue, but do not by themselves
    establish three-dimensional connectivity. Some narrow, approximately radial features in panel (b)
    correspond to known globular clusters
    (e.g. NGC~6121 and NGC~6397), with elongation predominantly
    along the line of sight, consistent with parallax-induced
    distance broadening.}
    \label{fig:all_sky_snake}
\end{figure*}

Figure~\ref{fig:all_sky_snake} shows the catalogue member stars in Galactic coordinates and in the heliocentric Cartesian \(X\)--\(Z\) plane.\footnote{https://astrotian.github.io/snake/visualization/} The distribution is strongly concentrated towards the Galactic plane, as expected for Galactic-disk stellar populations, and also shows projected substructure away from the highest-density plane regions. This figure provides a global view of the
sky coverage and vertical extent of the final catalogue. A more detailed interpretation of the spatial distribution and its connection to large-scale Galactic-disk structure is given in Section~\ref{subsec:spatial_distribution_spiral_arm}.

\subsection{PointNet Parameter Validation}
\label{subsec:model_validation}

After deriving homogeneous node parameters with the PointNet regressor,
we assess their accuracy before using them in the final catalogue
and subsequent catalogue-level analyses.
As a controlled test, we evaluate the catalogue estimator on
20,000 synthetic populations not used to train its weights,
using the mean of
100 predictions per population as described in Section~\ref{sec:pointnet}.
The test-set mean absolute errors (MAEs) are
\(0.079\) dex for metallicity \([{\rm Fe/H}]\),
\(0.058\) dex for logarithmic age,
\(0.066\) mag for visual extinction \(A_V\),
\(0.018\) for the photometric purity proxy \(f_{\rm cl}\), and
\(0.024\) mag for distance modulus \(DM\).
The empirical broadening parameter is recovered with an MAE of
\(1.62\). These synthetic tests show that the model recovers the input parameters within the simulated-data domain, while its behaviour on real observations is assessed through the external comparisons below.
\add{Using residual standard deviations in both domains,
the real-to-synthetic scatter ratios are 2.79 for Hunt ages,
2.57 for Hunt extinctions, and 1.43 and 1.57 for the GALAH
and clean Gaia RVS metallicity comparisons, respectively.} We therefore distinguish synthetic-test
accuracy, internal Monte-Carlo precision, and external agreement,
without applying a common uncertainty-inflation factor.

We then assess the model on real observational data by comparing the
predicted cluster parameters with two literature catalogues. We use the catalogue of \citet{hunt2023improving}, based on CNN modelling of two-dimensional cluster representations, and the catalogue of \citet{cavallo2024parameter}, based on an Artificial Neural Network combined with a QuadTree-based spatial feature extractor. These two catalogues provide external benchmarks for
overlapping samples of known open clusters, thereby reducing
sample-selection differences in the comparison.

\begin{figure*}[htpb]
    \centering
    \includegraphics[width=0.8\textwidth]{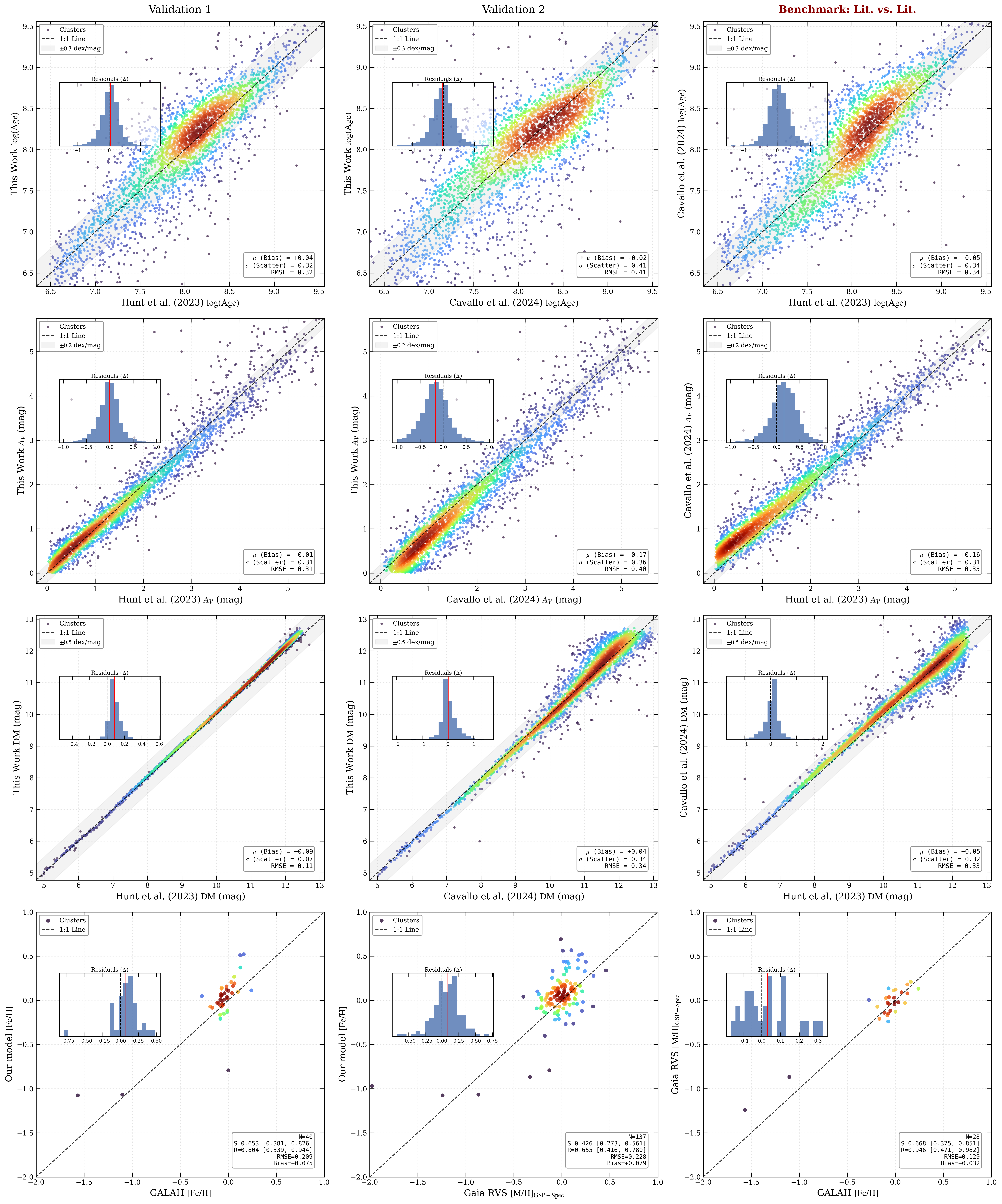}
    \caption{Combined external comparison of the PointNet-derived
parameters for the benchmark stellar systems. The first three rows compare logarithmic age, visual extinction, and distance modulus, respectively, against the literature catalogues of \citet{hunt2023improving} and \citet{cavallo2024parameter}. The left and middle columns show the comparisons between this work and the two external catalogues, while the right column shows the benchmark comparison between the two literature catalogues themselves. Dashed lines indicate the 1:1 relation, and the grey bands in the first three rows mark tolerance intervals of \(\pm 0.3\) dex in \(\log t\), \(\pm 0.2\) mag in \(A_V\), and \(\pm 0.5\) mag in \(DM\). Inset histograms show the residual distributions, with the red vertical line marking the mean offset. The fourth row provides an external spectroscopic consistency check for the predicted cluster metallicities. The first metallicity panel compares the model-predicted \([{\rm Fe/H}]\) with GALAH \([{\rm Fe/H}]\); the second compares the model-predicted \([{\rm Fe/H}]\) with the clean Gaia RVS GSP-Spec global metallicity, \([{\rm M/H}]_{\rm GSP\mbox{-}Spec}\); and the third compares GALAH \([{\rm Fe/H}]\) with Gaia RVS \([{\rm M/H}]_{\rm GSP\mbox{-}Spec}\) on their common cluster subset. The \(DM\) panels assess the consistency of the parallax-anchored
distance scale. These comparisons are used as external consistency
checks rather than as hard selection criteria for the final
Stellar Snake catalogue.}
    \label{fig:pointnet_validation}
\end{figure*}

The first three rows of Figure~\ref{fig:pointnet_validation} summarise
the external comparisons for logarithmic age, visual extinction, and
distance modulus. Relative to \citet{hunt2023improving}, the PointNet estimates show small mean offsets of \(+0.04\) dex in \(\log t\), \(-0.01\) mag in \(A_V\), and \(+0.09\) mag in \(DM\), with corresponding scatters of \(0.32\) dex, \(0.31\) mag, and \(0.07\) mag. The comparison between the two literature catalogues themselves provides a useful reference for the level of catalogue-to-catalogue systematics: Cavallo versus Hunt gives scatters of \(0.34\) dex in \(\log t\), \(0.31\) mag in \(A_V\), and \(0.32\) mag in \(DM\). The age and extinction estimates are broadly consistent with the
external catalogue scales. The substantially smaller \(DM\) scatter,
however, reflects the shared parallax dependence of the production output
and the external catalogues, and is therefore interpreted only as a
consistency check of the parallax-anchored astrometric distance scale.
For a CMD-only ablation excluding all per-star and node-level
parallax information, the residual standard deviation and MAE,
averaged over three training runs, are \(0.613\) and \(0.667\) mag
relative to \citet{hunt2023improving}, and \(0.673\) and \(0.714\) mag
relative to \citet{cavallo2024parameter}, respectively.

Metallicity is more difficult to infer from photometric and astrometric point clouds alone, because the CMD morphology is affected by degeneracies among age, distance, extinction, binarity, field contamination, and the adopted stellar-evolution models. We therefore treat the PointNet metallicity primarily as an auxiliary reference parameter rather than as a hard reliability criterion. To test whether the predicted metallicities retain physically meaningful information, we compare them with external spectroscopic metallicity scales. We adopt GALAH DR4 \citep{buder2025galah} as the primary high-resolution spectroscopic benchmark. At the stellar level, we require clean spectroscopic and abundance flags and retain only stars with \(T_{\rm eff} \ge 4500\,\mathrm{K}\). At the system level, we require at least five quality-filtered
spectroscopic members, yielding 40 GALAH systems: 32 open clusters,
six moving groups, and two globular clusters, NGC~6544 and NGC~6121. As an independent auxiliary benchmark, we also construct a clean Gaia RVS
comparison sample from the GSP-Spec global metallicity
\texttt{mh\_gspspec} \citep{recio2023gaia}, denoted here as
\([{\rm M/H}]_{\rm GSP\mbox{-}Spec}\).
We use the released GSP-Spec values without applying an additional
surface-gravity-dependent metallicity calibration.
We apply the same minimum-member requirement together with strict
\texttt{flags\_gspspec} cleaning, yielding 137 clusters. Since no
\([{\rm M/H}] + [{\rm Fe/M}]\) conversion is applied, the RVS comparison
should be interpreted as a metallicity-scale consistency check rather than
a strict \([{\rm Fe/H}]\)-to-\([{\rm Fe/H}]\) validation.

The fourth row of Figure~\ref{fig:pointnet_validation} shows the spectroscopic metallicity comparison. We focus primarily on the Spearman coefficient to assess metallicity
rank-order consistency, while reporting Pearson correlation and RMSE
as complementary measures of linear association and absolute agreement. Against GALAH \([{\rm Fe/H}]\), the PointNet metallicities show a Spearman rank coefficient of \(S = 0.653^{+0.173}_{-0.272}\), a Pearson coefficient of \(R = 0.804^{+0.140}_{-0.465}\), an RMSE of \(0.209\,\mathrm{dex}\), and a mean offset of \(+0.075\,\mathrm{dex}\). Excluding NGC~6544 and NGC~6121 leaves a positive rank correlation
of \(S=0.595\), compared with \(0.653\) for the full sample, whereas
Pearson \(R\) decreases from \(0.804\) to \(0.451\).
The RMSE is \(0.199\,\mathrm{dex}\) for the remaining \(N=38\) systems.
Thus, the rank association persists beyond the two metal-poor systems,
although the full-sample linear correlation is sensitive to their inclusion. Against the clean Gaia RVS \([{\rm M/H}]_{\rm GSP\mbox{-}Spec}\) scale, the agreement is weaker but remains positive, with \(S = 0.426^{+0.135}_{-0.153}\), \(R = 0.655^{+0.125}_{-0.239}\), \(\mathrm{RMSE} = 0.228\,\mathrm{dex}\), and a mean offset of \(+0.079\,\mathrm{dex}\). For context, the direct GALAH--clean-RVS overlap gives \(S = 0.668^{+0.184}_{-0.293}\), \(R = 0.946^{+0.036}_{-0.475}\), \(\mathrm{RMSE} = 0.129\,\mathrm{dex}\), and a mean offset of \(+0.032\,\mathrm{dex}\). Taken together, the GALAH sensitivity test and the weaker clean
GSP-Spec comparison support only limited external rank-order consistency
for the PointNet metallicities. We therefore report the predicted metallicities in the released catalogue as auxiliary reference parameters, but do not use them as hard criteria for physical association or include them in the fiducial SRI.

\subsection{Global Properties of the Stellar Snake Catalogue}
\label{subsec:global_properties}

\begin{figure}[htpb]
    \centering
    \includegraphics[width=0.45\textwidth]{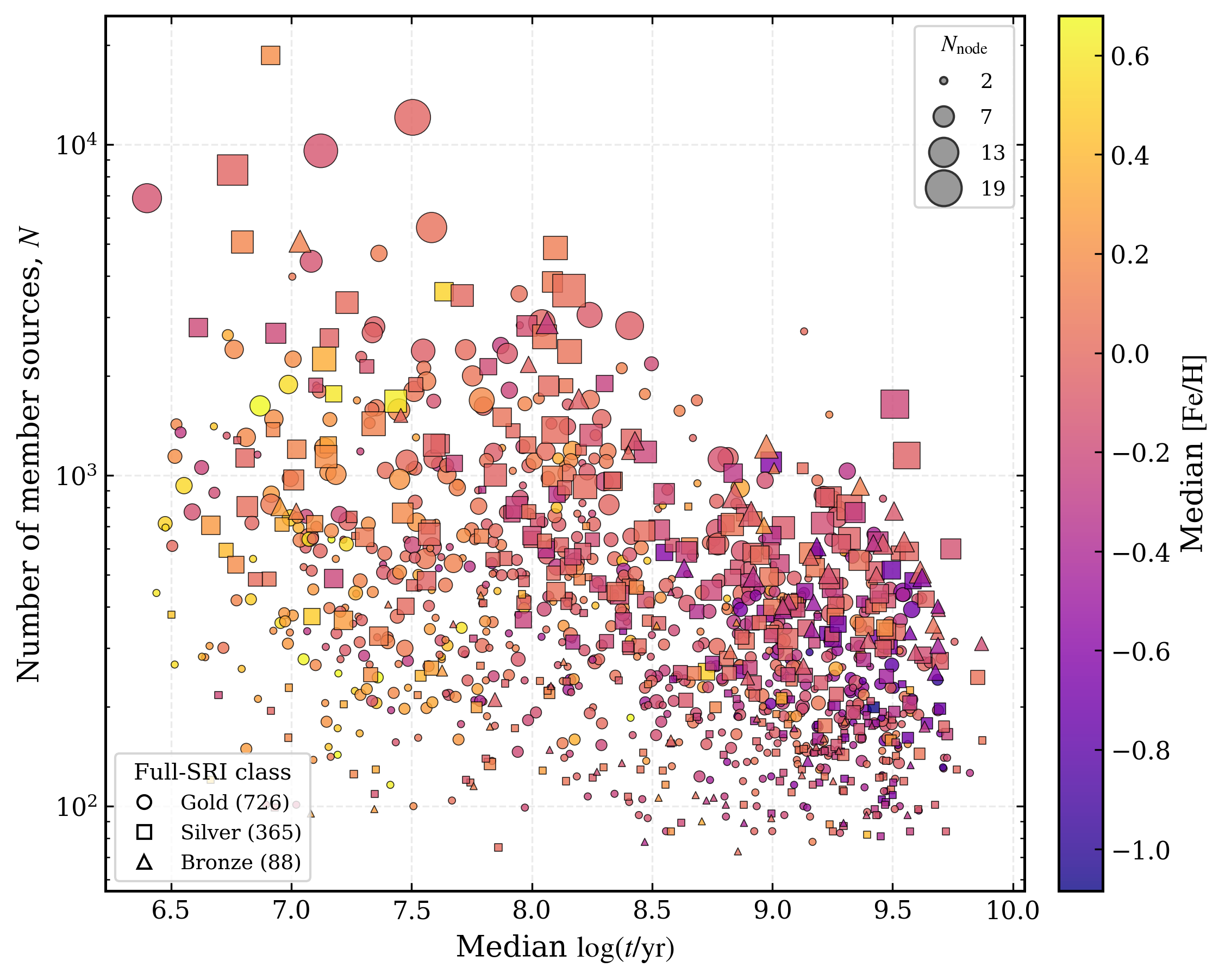}
    \caption{Global demographic properties of the SRI-scored Stellar Snake sample. The scatter plot shows the total number of member-star entries \(N\), on a logarithmic scale, as a function of the median logarithmic age \(\log t\) of each macro-structure. Point colour indicates the median PointNet-inferred metallicity \([{\rm Fe/H}]\), marker size is proportional to the number of constituent base nodes, \(N_{\rm node}\), and marker shape denotes the final SRI quality class. The distribution shows a broad age--metallicity pattern and a declining upper envelope in member-star entries toward older ages. Age and metallicity medians use base-node member counts as weights.
    The age--metallicity pattern is descriptive, not evidence for
    Galactic-disk chemical evolution; see
    Appendix~\ref{app:pointnet_residuals} for the residual-correlation test.}
    \label{fig:evolution}
\end{figure}

We next examine the global demographic properties of the final Stellar Snake catalogue. Figure~\ref{fig:evolution} shows the total number of member-star entries, \(N\), as a function of the median logarithmic age, \(\log t\), for Snakes with finite inferred age and metallicity. The colour of each point encodes the median PointNet-inferred metallicity, \([{\rm Fe/H}]\), the marker size represents the number of constituent base nodes, \(N_{\rm node}\), and the marker shape indicates the final SRI quality class. The age and metallicity medians are computed over the constituent
base nodes, weighted by their member counts
(Table~\ref{table:snake_complex}).

Two broad trends are visible in this parameter space. First, the sample shows an age--metallicity pattern in which younger Snake candidates tend to occupy the more metal-rich part of the diagram, whereas older systems are more frequently found at lower inferred metallicity. In the synthetic evaluation sample, the true and predicted
age--metallicity correlations are both close to zero:
\(R_{\rm true}=0.0118\) and \(R_{\rm pred}=0.0022\), with
corresponding Spearman coefficients of \(S_{\rm true}=0.0118\)
and \(S_{\rm pred}=0.0019\). Although the signed prediction
residuals are positively correlated
(\(R=0.3391\), \(S=0.2543\);
Appendix~\ref{app:pointnet_residuals}), this coupling does not
produce an appreciable global linear or monotonic
age--metallicity relation across the synthetic predictions.
Its positive sign also differs from the negative pattern
shown in Figure~\ref{fig:evolution}; the detected residual
coupling alone therefore does not establish that the catalogue
pattern is a regression artefact. These tests do not exclude
conditional biases or a synthetic-to-real domain gap.
We retain the pattern as a descriptive feature of jointly
inferred parameters, rather than as independent evidence
for Galactic-disk chemical evolution.

Second, the upper envelope of the observed member-star distribution
decreases toward older inferred ages. Young Stellar Snake candidates can contain large numbers of member-star
entries and multiple constituent base nodes. At older ages, systems with more than \(10^3\) member-star entries become less common, and the number of retained constituent nodes is generally smaller.

The negative detected-count--age association persists among
311 Gold systems with assessed \({\rm KPD}=0\), divided into four
approximately equal-count distance strata: their Spearman coefficients
range from \(-0.665\) to \(-0.451\). Upper-quantile regressions of \(\log_{10}N\) on \(\log t\),
with separate distance-stratum intercepts, also give negative
slopes at \(\tau=0.80,0.90,\) and \(0.95\)
(\(-0.344,-0.344,\) and \(-0.420\), respectively).
This is a control for distance and detected subdivisions, not a
completeness correction. The observed decline may reflect both dynamical
evolution and age-dependent detectability; their contributions cannot
be separated without an end-to-end selection-function calibration.

\subsection{Spatial Association with Spiral Arms}
\label{subsec:spatial_distribution_spiral_arm}

\begin{figure}[!htbp]
    \centering
    \includegraphics[width=\columnwidth]{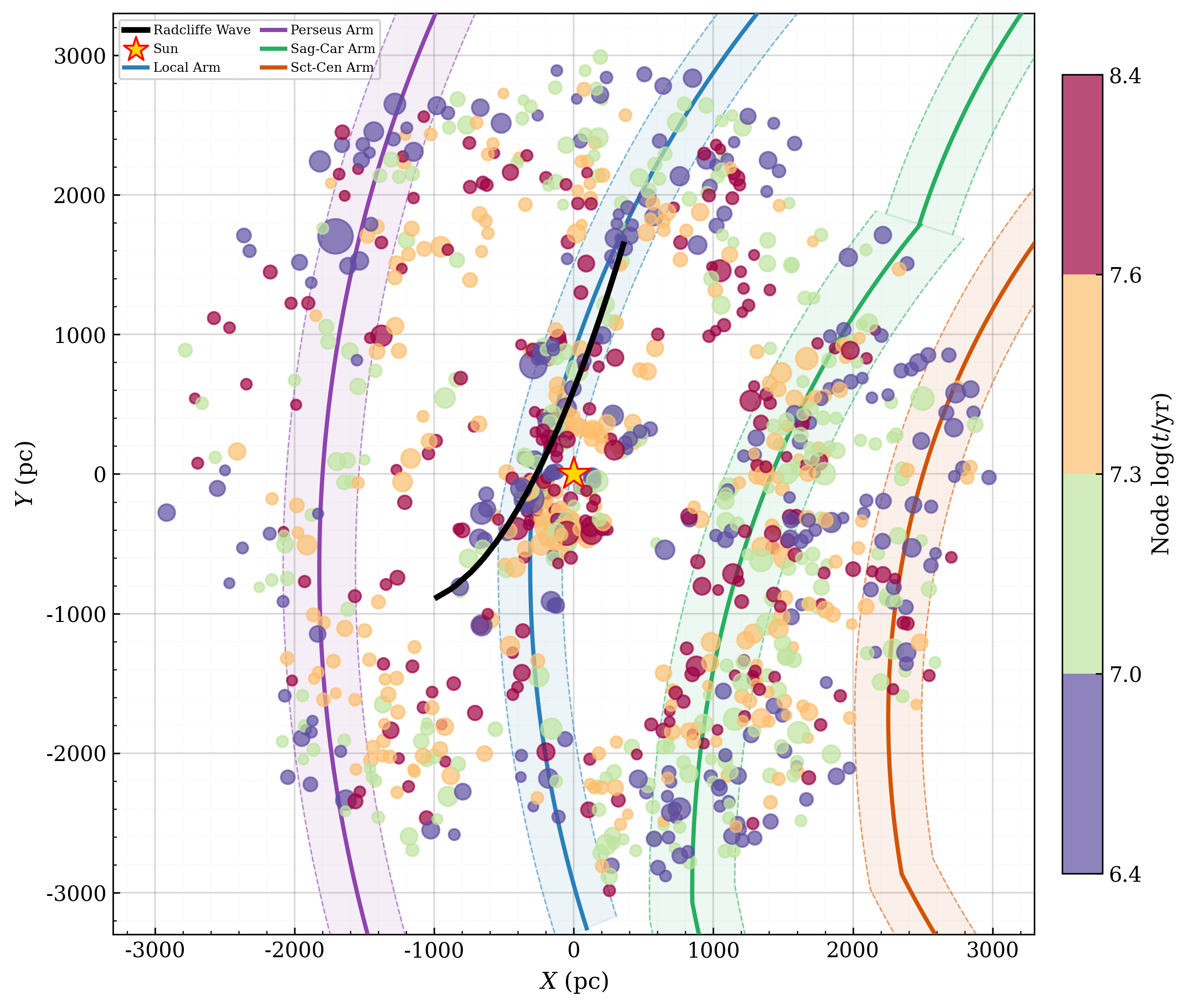}
    \caption{
    Heliocentric Galactic-plane \(X\)--\(Y\) projection of constituent nodes
    belonging to young Stellar Snake candidates with
    median logarithmic age, weighted by base-node member counts,
    \(\log(t/\mathrm{yr}) \leq 7.7\).
    Only nodes within 3~kpc are shown.
    Each point represents one Snake node. The marker size
    increases with the number of member stars in the node,
    and the colour encodes the individual PointNet-inferred node age.
    The coloured solid curves and their shaded or dashed envelopes show
    the spiral-arm loci and approximate widths from the combined-tracer
    model of \citet{hou2021spiral}.
    The black curve marks the Radcliffe Wave from \citet{alves2020galactic},
    and the star symbol marks the position of the Sun.
    }
    \label{fig:young_snake_spiral}
\end{figure}

To examine whether the young Stellar Snake population is spatially associated with large-scale Galactic star-forming structures, we project the constituent nodes of young Snake candidates onto the heliocentric Galactic-plane \(X\)--\(Y\) plane in Figure~\ref{fig:young_snake_spiral}. We use the individual nodes, rather than the mean centres of the full Snake candidates, as the plotting units. This choice preserves the internal filamentary skeleton of the extended complexes and avoids artificially compressing multi-node systems into single centroids.

For this comparison, we restrict the displayed sample to
nodes within 3~kpc belonging to Snakes whose median logarithmic
age, weighted by base-node member counts, satisfies
\(\log(t/\mathrm{yr}) \leq 7.7\), corresponding to approximately 50 Myr. The marker size increases with the number of member stars in each node,
while the colour encodes the individual node age inferred by the PointNet regressor. The spiral-arm loci and their approximate widths are taken from the combined-tracer model of \citet{hou2021spiral}, which was constructed using multiple young and star-forming tracers, including GMCs, HMSFR masers, HII regions, O-type stars, and young open clusters. We adopt this combined-tracer model, rather than relying only on maser-based arm loci, because it provides a more continuous description of nearby spiral-arm segments within the solar-neighbourhood volume considered here. The Radcliffe Wave is overlaid using the spatial trace reported by \citet{alves2020galactic}, providing an independent local gaseous-filament reference. The overlaid loci and envelopes provide spatial references,
not criteria for assigning nodes to individual arms.

The resulting distribution shows a projected spatial correspondence
between young Snake nodes and nearby large-scale star-forming structures.
In the solar vicinity, a prominent concentration of young nodes
lies near the projected Radcliffe Wave trace.
Additional concentrations are found along the Local, Perseus,
Sagittarius--Carina, and Scutum--Centaurus arm loci.
In particular, the group of young nodes near
\(X \simeq 1\,\mathrm{kpc}\) and \(Y \simeq -2.5\,\mathrm{kpc}\)
falls within the projected Sagittarius--Carina arm band of the
Hou combined-tracer model and corresponds to the near-side
Carina-tangent sector.

Nodes with older inferred ages can remain in the young-Snake sample
because the selection uses each Snake's weighted median age rather
than an age cut on every node.
\add{We compared distances to the arm centrelines with Hunt--Reffert
clusters matched in age, heliocentric distance, and \(|Z|\) within
the same nearest-arm region, excluding clusters that share members
with the Snake catalogue. The median Snake-level difference
(Snake minus control) is \(+31\)~pc, with a 95\% bootstrap interval
of \([-34,+58]\)~pc, providing no evidence of excess arm proximity.}

\add{For the Radcliffe Wave, selection within 135~pc of its
projected trace in the \(X\)--\(Y\) plane, without using \(Z\),
yields 112 nodes in 29 Snakes. Comparing per-Snake median observed
heights with the corresponding Wave heights gives
Spearman \(\rho=0.665\) (\(p_{\rm perm}\simeq10^{-4}\)).
This supports a vertical morphological correspondence in the
selected sample.}

\section{Discussion}
\label{sec:discussion}

In this section, we assess the reliability and broader context of the final Stellar Snake catalogue through three complementary lines of evidence. We first compare our results with external open-cluster system catalogues to evaluate consistency and complementarity (Section~\ref{subsec:external_oc_systems}). We then examine the photometric-purity proxy using radial-velocity
scatter and assess inter-node chemical consistency against matched
real-node controls (Section~\ref{sec:node_coherence}). Finally, we revisit the original Stellar Snake region from Paper I to illustrate how this source-level-to-complex framework represents the region (Section~\ref{subsec:re-evaluating_snake}).

\subsection{External Comparisons with Open-Cluster Systems}
\label{subsec:external_oc_systems}

To further assess the reliability and scope of the Stellar Snake catalogue, we compare our results with recent open-cluster (OC) system catalogues. These external catalogues provide useful reference samples because they are constructed from independently identified OCs and apply different grouping criteria. At the same time, the comparison must be interpreted carefully: OC-system catalogues and our Snake catalogue do not define structures at exactly the same hierarchy level. The former usually start from known open clusters, while our method starts from member-star-level FoF overdensities and then links statistically significant nodes into larger complexes. The following comparison is therefore intended to test external consistency and catalogue complementarity, rather than to provide a one-to-one replacement for previous OC-system catalogues.

\subsubsection{Catalogue-level Cross-analysis with OC Systems}
\label{subsubsec:palma_oc_systems}

As an external benchmark at a different hierarchy level, we compare the final
Snake catalogue with the paired and multiple open-cluster systems of
\citet{palma2025binary}, which are built on the Hunt--Reffert cluster sample.
The comparison is not a one-to-one catalogue replacement test: Palma systems
link catalogued open clusters, whereas a Snake links statistically significant
member-star overdensities that need not be classical open clusters. We
therefore use the Palma classifications as external, noisy physical labels and
ask whether the reported SRI responds to these labels when the Palma systems
are rebuilt as pseudo-Snakes.

We first define the cluster-entry sample used for the external OC-system
comparison. For every Palma HR24 cluster entry, we retain Hunt members with
parallax signal-to-noise ratio \(\varpi/\sigma_\varpi>10\), heliocentric
distance \(d<3.1\) kpc, and at least 30 retained members. This yields
1{,}140 comparable Palma/Hunt cluster entries. At the FoF-hierarchy level,
991 of these entries (86.9\%) have a corresponding source-level overdensity
before final Snake assembly, indicating that the comparison sample is largely
represented in our FoF hierarchy at the open-cluster-entry level. This check
defines the cluster-level denominator for the Palma comparison; it is not used
as a completeness estimate for the final Snake catalogue or for the
member-star catalogue.

For the SRI consistency comparison, we apply an additional membership-probability cut when
constructing the pseudo-Snake member clouds. Each Palma system is represented
as a pseudo-Snake at the component level: each retained HR24 cluster is first
represented as one pseudo-base component, and its Hunt members with membership
probability \(>70\%\), \(\varpi/\sigma_\varpi>10\), and \(d<3.1\) kpc define
the corresponding member cloud. Only HR24 clusters with at least five members
after these cuts are retained. The same SRI entity-contraction and scoring
framework is then applied before reporting the SRI, so the final number of
graph entities may be smaller than the number of retained HR24 clusters. We
require each pseudo-Snake to contain at least two retained cluster components
before scoring. Sources shared between cluster components are assigned to the
component with the higher membership probability; systems with
unresolved equal-probability assignments are excluded.
The component-size requirement is checked again after this assignment.
The resulting sample contains 533 paired systems:
48 genetic binaries (B), 78 tidal-capture pairs (C),
44 coeval optical pairs (Oa), and 363 optical pairs (O).
A further 250 systems are retained from the multiple-system sample.
This designation follows the parent catalogue; some systems retain
only two components after filtering.
No background-star support is added to the Palma member clouds. The SRI calculation is performed without using the
Palma class labels, ages, or colour--magnitude information. \add{The Palma benchmark informed assessment of scoring choices
and recommended thresholds, so it is not a held-out validation sample.}
The comparison is also not information-independent:
the Palma classifications use proper-motion distributions together with
age and colour--magnitude information, while the SRI also contains
tangential-velocity coherence. We therefore interpret the Palma comparison
as an external consistency diagnostic rather than as an independent
validation. Moreover, because the labelled Palma systems probe neighbouring
clusters within approximately 50~pc, this comparison primarily tests the
nearby pair regime rather than the linking of extended macro-structures.

\begin{figure*}[t]
    \centering
    \includegraphics[width=\textwidth]{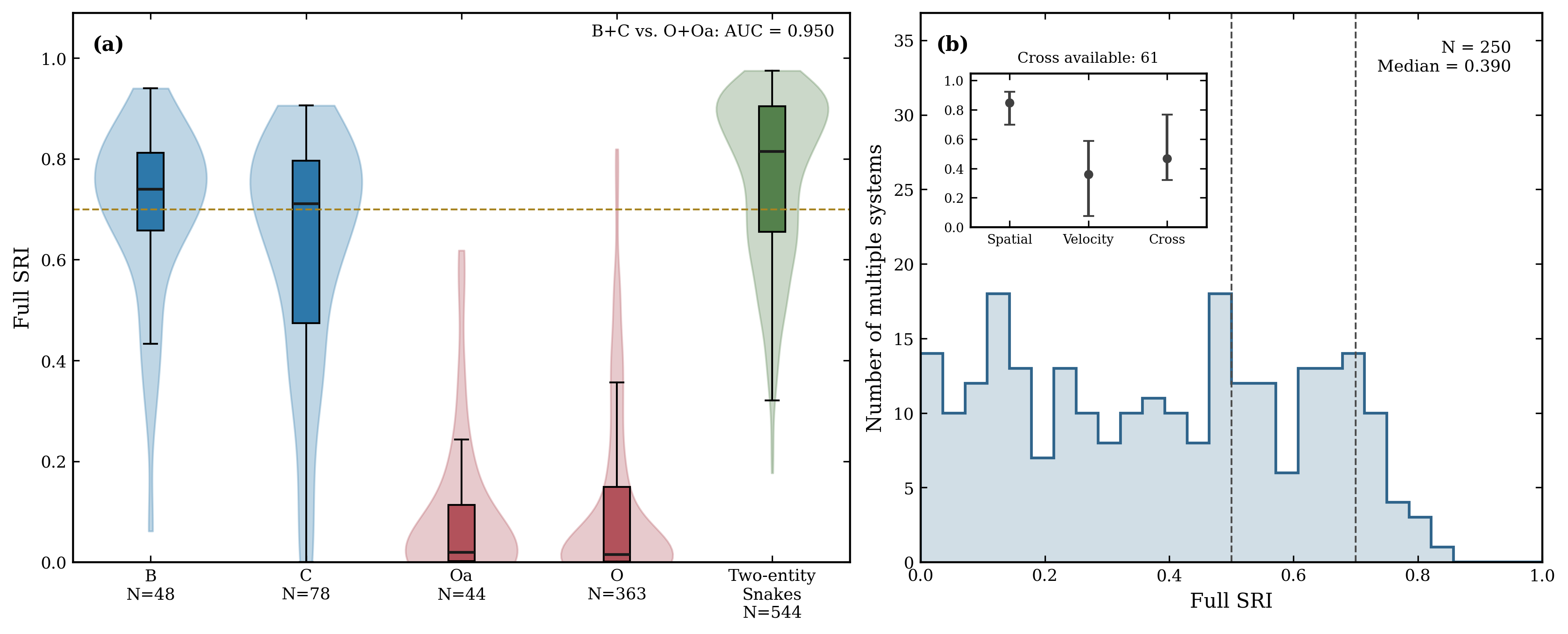}
    \caption{External consistency of the continuous full SRI with
    the open-cluster systems of \citet{palma2025binary}.
    Filtered Hunt--Reffert cluster member clouds define pseudo-base
    components, followed by the same entity contraction and SRI calculation,
    without companion-star support. (a) The 533 paired systems:
    48 B, 78 C, 44 Oa, and 363 O, compared with 544 two-scoring-entity
    Snakes. The B+C versus O+Oa AUC is 0.950; the dashed line marks
    the recommended Gold boundary. Palma class labels do not denote
    SRI evidence groups. (b) The 250 multiple systems, with median
    full SRI 0.390. Dashed lines mark the 0.50 and 0.70 boundaries.
    The inset shows pillar-score medians and 16th--84th percentiles;
    Cross uses only the 61 systems for which it is active.
    The comparison is label-blind but shares proper-motion information
    with the Palma classifications.}
    \label{fig:palma_sri_validation}
\end{figure*}

The 533 paired systems provide a labelled consistency benchmark
(Figure~\ref{fig:palma_sri_validation}a). For the 126 B+C pairs and
407 O+Oa optical controls, the median full SRI values are 0.728 and
0.016, respectively; \({\rm AUC}=0.950\), and the optical-control
95th percentile is 0.512. At the recommended Gold boundary of 0.70,
69/126 B+C and 7/407 optical pairs are selected, giving B+C recall
54.76\%, optical-control false-positive fraction 1.72\%, and
benchmark precision 90.79\%. At 0.50 these quantities are
79.37\%, 5.41\%, and 81.97\%; at 0.80 they are
25.40\%, 0.74\%, and 91.43\%, respectively.
Precision depends on the class mixture of this benchmark, and recall
refers to these labelled pairs, not catalogue-wide purity or completeness.
All scored Palma pairs belong to SRI evidence group A; Palma's B/C
physical labels are unrelated to the SRI A/B/C evidence groups.

As an internal comparison, the 544 catalogue Snakes represented
by two scoring entities have a median SRI of 0.815.
This selection refers to the entities used in scoring, rather than
a separate sample of naturally uncontracted pairs.
The comparison is descriptive: the catalogue scores retain their
adopted spatial bridge support, whereas the Palma systems are scored
from cluster members alone. The separation from optical pairs supports
external consistency within this benchmark, but does not establish
the physical nature of every individual association.

The 250 retained multiple systems provide an unlabelled
comparison (Figure~\ref{fig:palma_sri_validation}b).
Their median full SRI is 0.390, with 25 Gold, 63 Silver, and
162 Bronze systems. The spatial and velocity pillar medians are
0.847 and 0.358. Cross is active for 61 systems, with median 0.467;
this conditional median is not a whole-sample statistic.
These systems illustrate score behaviour beyond simple pairs, but do
not provide supervised physical--optical labels or a calibration of
higher-order association purity.

The Palma comparison is therefore a nearby-system consistency
check with shared proper-motion information, not an
information-independent validation or a macro-structure fragmentation
measurement.

\subsubsection{Case-study Cross-analysis with OC Systems}
\label{subsubsec:liu_oc_groups}

The statistical comparison with \citet{palma2025binary} suggests that many
externally identified open-cluster systems are recovered by our catalogue and
placed in a broader stellar-complex context. We now examine this hierarchy in
a more concrete, case-by-case manner using the open-cluster groups reported
by \citet{liu2025formation}, who identified several inter-correlated
open-cluster groups (G1--G4) on the basis of spatial, kinematic, and age
coherence. These groups provide an independent, physically motivated
reference for assessing how the Snake catalogue represents known
open-cluster associations.

Among the four Liu et al.\ groups, G1, G3, and G4 each have a single dominant Snake counterpart, whereas G2 is the only case resolved into two major catalogue partitions. We focus on the G2 group as a representative example. All eight G2 cluster centres lie within 30~pc of a Snake base node, providing positional counterparts. In the present catalogue, however,
the G2 region is not assigned to a single Snake. Instead, the clusters are
split between two adjacent Snake labels. Six clusters (OCSN~16, OCSN~18,
Stephenson~1, UPK~101, UPK~78, and UPK~83) are matched to nodes belonging
to Snake~1101, whereas the two lowest-latitude clusters, UPK~64 and UPK~72
(\(l \approx 56\)--\(58^\circ\), \(b \approx 10^\circ\)), have positional counterparts in the neighbouring Snake~1102. This division is produced directly by the high-dimensional clustering. At the adopted catalogue resolution, the region is therefore represented by two adjacent substructures rather than being forced into a single extended label. These two partitions may still belong to a common higher-order complex.

\begin{figure*}[htbp]
    \centering
    \includegraphics[width=\textwidth]{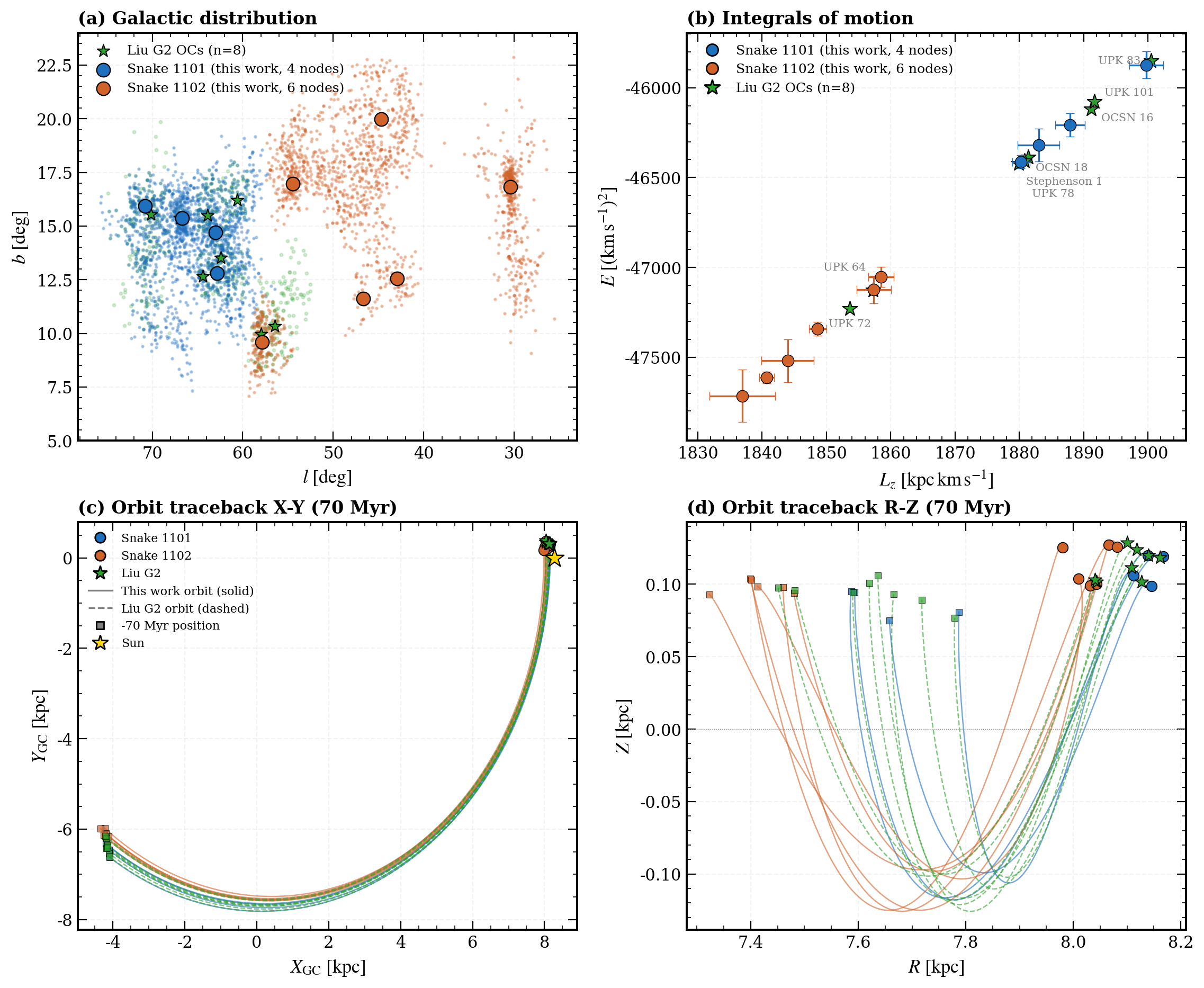}
    \caption{
    Dynamical consistency analysis between Snakes~1102/1101 from this work
    and the G2 group of \citet{liu2025formation}.
    \textbf{(a) Galactic distribution:} the six core G2 clusters are matched
    to Snake~1101 (blue), while the two lowest-latitude clusters, UPK~64 and
    UPK~72, have positional counterparts in the neighbouring Snake~1102 (orange); together they provide positional counterparts to all eight G2 clusters.
    \textbf{(b) Integrals of motion:} The two Snake partitions occupy adjacent but offset loci in \((L_Z,E)\); because these coordinates contribute to the clustering representation, this separation is shown as a conditional diagnostic rather than an independent validation. The
    \((L_Z,E)\) offset accompanies the partition despite the similar
    ages and distances of the two Snakes. Error bars on the Snake
    nodes denote the standard error of the node centroid in \((L_Z,E)\),
    estimated as \(0.741\times{\rm IQR}/\sqrt{N_{\rm member}}\).
    \textbf{(c \& d) Orbit traceback over 70 Myr:} the Galactocentric
    \(X\)--\(Y\) and meridional \(R\)--\(Z\) trajectories of both Snakes and
    the G2 clusters remain broadly parallel, consistent with the possibility that the two catalogue partitions belong to a broader higher-order association.
    }
    \label{fig:snake_liuG2}
\end{figure*}

The two Snakes are nevertheless closely related. As shown in
Figure~\ref{fig:snake_liuG2}, their node populations have similar ages
(\(\log t\simeq7.54\)--7.55), distances (approximately 350--365~pc),
and adjacent spatial distributions. After recomputing the Liu G2
RVs from member measurements with a robust median estimator, the two
catalogue partitions show an offset in \((L_Z,E)\) that persists under
delete-one-node tests. Because these quantities enter candidate linking,
the offset is a conditional consistency diagnostic, not independent
validation of the partition. The 70~Myr traceback shows broadly
parallel trajectories for the two Snakes and the G2 clusters in the
Galactocentric \(X\)--\(Y\) and \(R\)--\(Z\) projections.

This case illustrates a generic property of clustering in a feature space
that includes orbital integrals. A structure that is continuous in position,
distance, and age, but carries an internal kinematic gradient, can be
partitioned into adjacent labels when the gradient becomes significant in
\((E,L_Z)\). This is analogous to the relationship between different
\(\mathrm{id\_part}\) components within a single Snake: neighbouring
segments can share spatial and stellar-population properties while differing
in fine kinematic detail. The positional correspondence of UPK~64 and UPK~72 with
Snake~1102 rather than with the G2 core in Snake~1101 should therefore be read
as the clustering responding to a measurable kinematic gradient in the adopted
feature space, rather than as evidence that the two Snakes are physically
unrelated. We therefore interpret Snake~1101 and Snake~1102 more conservatively as two catalogue-level substructures within a possible larger hierarchical complex that overlaps the Liu G2 system. The partition is resolution dependent and may include fragmentation of a higher-order association; the present catalogue does not assign a physical fragmentation rate to this case.

\subsection{Kinematic and Chemical Diagnostics of Snake Base Nodes}
\label{sec:node_coherence}

\citet{zucker2022disconnecting} raised three interconnected objections to the stellar strings of \citet{kounkel2019untangling}: (i) apparent spatial coherence does not tighten with improved astrometry; (ii) member-level line-of-sight velocity dispersions are \(\sim 15\)--\(16\) km s\(^{-1}\), comparable to the virial velocity of a \(\sim 2 \times 10^{6} M_\odot\) bound system and far exceeding the observed cluster mass; and (iii) apparent chemical homogeneity can be reproduced by random ensembles of unrelated open clusters. The spatial and catalogue-reliability aspects are examined
through the FoF-topology consistency check, the SRI and separate
partition diagnostic (Section~\ref{sec:snake_reliability}), and
external open-cluster catalogue comparisons.
Here we use the RV comparison to assess the empirical behaviour
of the photometric-purity proxy and examine chemical consistency
through a separate node-level diagnostic.

The interpretation of these tests depends on the hierarchical nature of the Snake catalogue. A Snake is a relation among statistically significant base-node overdensities; it is not defined as a single bound object, and no assumption is made that every individual base node is dynamically relaxed. We therefore distinguish member-level scatter within a node from the systemic differences among nodes. The physically relevant question for the chemical comparison
is whether the abundances of nodes assigned to the same Snake
are more coherent than those of matched ensembles of real nodes
drawn from different Snakes.

For the kinematic diagnostic, we compare Snake base-node RV
 dispersions with a Hunt--Reffert cluster control, stratified by the
 photometric purity proxy \(f_{\rm cl}\).
 The chemical comparison uses count-matched and age--count-matched
 real-node ensembles, supplemented by the position--age-matched
 control described below.

\subsubsection{Kinematic coherence}
\label{subsubsec:kinematic-coherence}

A natural concern, raised for large-scale stellar streams by
\citet{zucker2022disconnecting}, is that a large member-level radial-velocity dispersion
would imply a short diffusion time if it were interpreted as the internal
velocity scale of a single coherent system. For the Snake catalogue, this quantity must be interpreted more carefully,
because the identified complexes are not claimed to be gravitationally
bound clusters and because raw member-level dispersions are sensitive
to measurement uncertainties, binaries, limited sampling, and
residual field contamination \citep{hunt2024improving}.

We therefore compare the Snake base nodes with the Hunt--Reffert open-cluster
catalogue using the same radial-velocity dispersion definition: the sample
standard deviation of member radial velocities, denoted \(\sigma_{V_R}\).
For the main comparison we require \(N_{\rm RV}\ge25\).
The same threshold is applied to the Hunt--Reffert sample.

We then examine \(\sigma_{V_R}\) as a function of the photometric purity
proxy \(f_{\rm cl}\). This quantity is predicted from
CMD morphology and richness, without explicit radial-velocity,
proper-motion, or parallax inputs.
Without a direct external purity reference for the real nodes,
this comparison provides an empirical check of the \(f_{\rm cl}\)
ranking.

\begin{figure}[htbp]
    \centering
    \includegraphics[width=\columnwidth]{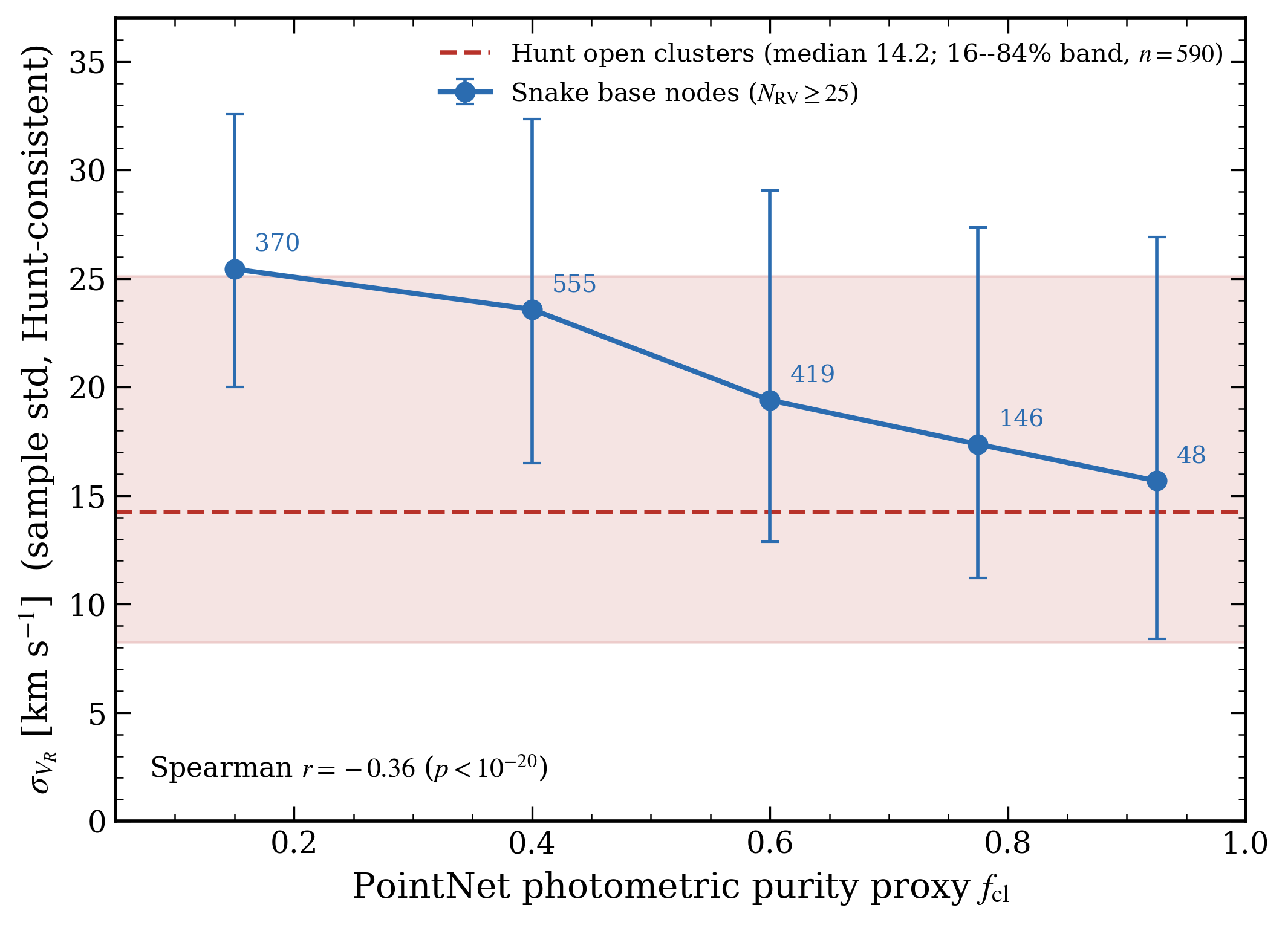}
    \caption{Raw member-level radial-velocity scale of Snake base nodes as a
function of the PointNet-inferred photometric purity proxy \(f_{\rm cl}\).
    The dispersion is measured with the same sample-standard-deviation
    definition as the Hunt--Reffert catalogue, and both samples require
    \(N_{\rm RV}\ge25\). Points show bin medians, error bars show 16--84th
    percentiles, and labels give the number of nodes per bin. The
    Hunt--Reffert open-cluster median is shown by the dashed line, with the
    shaded band marking its 16--84th percentile range. The monotonic decline toward higher \(f_{\rm cl}\)
(Spearman \(r=-0.36\), \(p<10^{-20}\)) provides an empirical real-data
consistency check on the purity ranking: lower-\(f_{\rm cl}\) nodes show
larger member-level RV scales, whereas high-\(f_{\rm cl}\) nodes approach
the Hunt catalogue-level scale measured with the same raw estimator.}
    \label{fig:kinematic_coherence}
\end{figure}

Figure~\ref{fig:kinematic_coherence} shows a clear decline in the raw
member-level RV scale with increasing \(f_{\rm cl}\) (Spearman \(r=-0.36\),
\(p<10^{-20}\)). Among the 1,538 nodes with
\(N_{\rm RV}\ge25\), the median raw scale is
\(22.34\,{\rm km\,s^{-1}}\); in the highest-\(f_{\rm cl}\) bin, containing
48 nodes, it decreases to \(15.68\,{\rm km\,s^{-1}}\). The 590 Hunt--Reffert open clusters satisfying the same
\(N_{\rm RV}\) requirement have a published median RV standard
deviation of \(14.24\,{\rm km\,s^{-1}}\). Thus, higher-\(f_{\rm cl}\) nodes
enter the same broad catalogue-level RV regime as the Hunt control, providing
an empirical real-data consistency check on the \(f_{\rm cl}\) ranking.

Because apparent magnitudes and richness can retain distance-dependent
information, we additionally calculated the partial Spearman correlation
between \(f_{\rm cl}\) and the raw member-level RV scatter while controlling
simultaneously for
\(d_\odot=(X^2+Y^2+Z^2)^{1/2}\) and the median \(G\) magnitude of the same
\(G<18\) member inputs. For the primary \(N_{\rm RV}\geq25\) sample
(\(N=1538\)), the coefficient changes from \(\rho=-0.361\) before the
controls to \(\rho_{\rm partial}=-0.206\) after the controls. The negative association persists after these controls, supporting
the empirical \(f_{\rm cl}\) ranking beyond the unadjusted correlation. We consequently interpret \(f_{\rm cl}\) as an empirical
ranking proxy for the observational cleanliness of the member sample,
rather than as a distance-independent or causal measure of true membership
purity.

As a measurement-error sensitivity test, we additionally fit a
single-component heteroscedastic Gaussian model,
\begin{equation}
v_i\sim\mathcal{N}\!\left(\mu,\sigma_{\rm eff}^{2}+\epsilon_i^{2}\right),
\end{equation}
where \(\epsilon_i\) is the reported stellar RV uncertainty. The median
\(\sigma_{\rm eff}\) is \(21.05\,{\rm km\,s^{-1}}\) for the full node
sample and \(12.68\,{\rm km\,s^{-1}}\) in the highest-\(f_{\rm cl}\) bin;
six of the 48 nodes in that bin have
\(\sigma_{\rm eff}\leq1.5\,{\rm km\,s^{-1}}\). Applying the same model to the available member RVs of the 590 Hunt
open clusters, with their membership cut \(P_{\rm mem}\ge0.5\),
gives a median effective scale of \(10.45\,{\rm km\,s^{-1}}\), and a robust MAD-based estimate shows the same qualitative
decrease toward higher \(f_{\rm cl}\). Because these estimators do not
include an explicit binary-orbit component, a field-star mixture, validated
star-by-star membership probabilities, or explicit outlier rejection, we
interpret \(\sigma_{\rm eff}\) as a measurement-error-deconvolved effective
RV scale rather than as the true intrinsic dispersion. These tests provide empirical support for the \(f_{\rm cl}\) ranking
and show that high-\(f_{\rm cl}\) nodes have catalogue-level RV scales
close to the Hunt control, while leaving intrinsic dispersions
undetermined.

\subsubsection{Chemical Coherence}
\label{sec:chemical_coherence}

Chemical abundances provide an independent but more sparsely sampled test of
the inter-node associations. We do not use chemistry as a catalogue-level
reliability criterion. Instead, following the chemical-coherence concern
raised for stellar strings by \citet{zucker2022disconnecting} and the
abundance analysis of \citet{manea2022galah}, we ask whether chemically
sampled base nodes in the same Snake are more coherent than matched
ensembles of real nodes drawn from other Snakes. This is a node-level
analogue tailored to the Snake catalogue, not a reproduction of the
open-cluster star-level random draw used by \citet{zucker2022disconnecting}.
It therefore tests relative inter-node coherence within the final Snake
catalogue while allowing chemically similar unrelated nodes to occur in the
control population.

We cross-match the Snake member catalogue with GALAH~DR4
\citep{buder2025galah}. We retain stars with \texttt{flag\_sp} = 0,
\texttt{flag\_fe\_h} = 0, zero-valued per-element abundance flags, and
\(T_{\rm eff} \geq 4500\) K, consistent with the quality control used for
the spectroscopic metallicity validation. Bridge stars, which by definition have no retained base-node assignment,
are excluded from the node-level chemical analysis. For each node and each of fifteen
GALAH abundance elements (O, Na, Mg, Al, Si, Ca, Ti, Cr, Mn, Fe, Ni, Cu, Zn,
Y, and Ba), we require at least five valid spectroscopic stars per element
and at least five valid elements per node, and adopt the median stellar
abundance as the node abundance.

The chemical sample contains 60 chemically qualified nodes
distributed over 42 Snakes. The within-Snake dispersion can be evaluated
for the 30 nodes belonging to the 12 Snakes with at least two chemically
qualified nodes. The remaining 30 qualified nodes belong to 30 distinct Snakes
with one chemically qualified node each and provide the candidate
control pool for the strict local matching test described below. This limited coverage reflects the
GALAH footprint, magnitude limit, and evolutionary-stage selection, so the
chemical results characterise the chemically sampled subset rather than
the catalogue as a whole.

For each element, we estimate the within-Snake dispersion of node abundances
with a robust MAD statistic and divide it by the median dispersion of matched
random ensembles. Values below unity therefore indicate enhanced chemical
coherence among nodes in the same Snake. We use two catalogue-wide control ensembles.
The count-matched null preserves the observed number of chemically sampled
nodes in each Snake and draws chemically valid nodes from other Snakes in the
same final base-node catalogue. The age- and count-matched null additionally
replaces each real node with a node of similar model-inferred age from
another Snake, using \(\lvert \Delta \log t \rvert \leq 0.25\) dex with a
fallback to twice this tolerance when necessary. Because the node ages are
model inferred and the chemically sampled subset is small, the age-matched
null is treated as a conservative robustness check rather than as an exact
physical conditioning variable. 

The median within-Snake-to-control ratios across the fifteen elements
are 0.40 and 0.51 on the \([X/H]\) scale for
the count-matched and age-and-count-matched controls, respectively. On the
diagnostic \([X/{\rm Fe}]\) scale, the corresponding ratios are 0.52 and
0.78. The catalogue-wide comparisons therefore retain a broad directional
contrast toward lower abundance dispersion among same-Snake nodes.
However, because these global controls do not fully condition on the local
spatial environment, they are treated as descriptive comparisons rather
than as a position-matched test of common chemical origin.

To control the local spatial environment more directly, we also
constructed an abundance-blind, non-reuse matched control using
pre-specified common-support criteria in group-centre position, spatial
scale, mean age, and GALAH quality-member count. The hard requirements
were a group-centre separation of at most 300~pc, a control-to-target
RMS-radius ratio between 0.5 and 2.0, a mean-\(\log t\) difference of at
most 0.25~dex, and a control-to-target ratio between 0.5 and 2.0 in the median
per-node number of GALAH sources passing the basic quality cuts. Control nodes and control Snakes were not reused.

Under these criteria, only two of the twelve chemically evaluable
Snakes could be matched. On the \([X/H]\) scale, 14 of the 15 element-wise
dispersion ratios are below unity and the median-across-elements ratio is
\(R_{\rm chem}=0.2412\). However, two independent matched blocks permit only \(2^2=4\)
target--control label assignments, applied jointly across all elements to preserve their dependence, giving exact one-sided and two-sided
probabilities of \(p=0.25\) and \(p=0.50\), respectively,
with the two-sided test based on \(|\ln R_{\rm chem}|\). The present
GALAH footprint therefore preserves the same directional chemical contrast
but does not provide sufficient local common support for a statistically
powered position--age-matched test.

As a secondary, pair-level diagnostic, we also compute a multi-element
chemical-doppelganger distance between node pairs. For two nodes \(i\) and
\(j\), we define
\begin{equation}
D^2 =
\frac{1}{N_e}
\sum_e
\frac{(m_{i,e} - m_{j,e})^2}{s_{i,e}^2 + s_{j,e}^2},
\label{eq:chem_d2}
\end{equation}
where \(m_{i,e}\) and \(s_{i,e}\) are the median abundance and statistical
uncertainty of node \(i\) for element \(e\), and the sum is taken over the
\(N_e\) elements valid for both nodes, requiring at least five shared
elements. For nodes with identical underlying abundance means, independent
Gaussian measurement errors, and correctly calibrated uncertainties,
the expectation is \(\mathbb{E}[D^2]=1\). Intrinsic differences between
nodes would add to this expectation. In
practice, the abundance dimensions are correlated and the node pairs are not
statistically independent, so we use \(D^2\) only as an intuitive relative
distance against matched real-node controls rather than as a formal
\(\chi^2\) probability.

For the 24 same-Snake node pairs in the chemically sampled subset, the
median \([X/H]\) distance is \(D^2=1.82\), whereas age-matched cross-Snake
node-pair draws have a median \(D^2=5.10\). The corresponding descriptive
node-pair-level chemical-doppelganger frequency, defined as the fraction of
matched cross-Snake draws with \(D^2\) no larger than the median same-Snake
value, is 14.3\%. On the \([X/\mathrm{Fe}]\) scale, the corresponding medians
are \(D^2=4.15\) and \(9.31\), with a doppelganger frequency of 17.1\%.
Same-Snake pairs are shifted toward smaller chemical distances, but the
overlap with the cross-Snake control remains non-negligible. These frequencies
are substantially higher than the order-\(1\%\) star-level field rate reported
by \citet{ness2018galactic}, although the two statistics are not directly
equivalent because our comparison is based on node-level abundance summaries.

\begin{figure*}[htbp]
    \centering
    \includegraphics[width=0.95\textwidth]{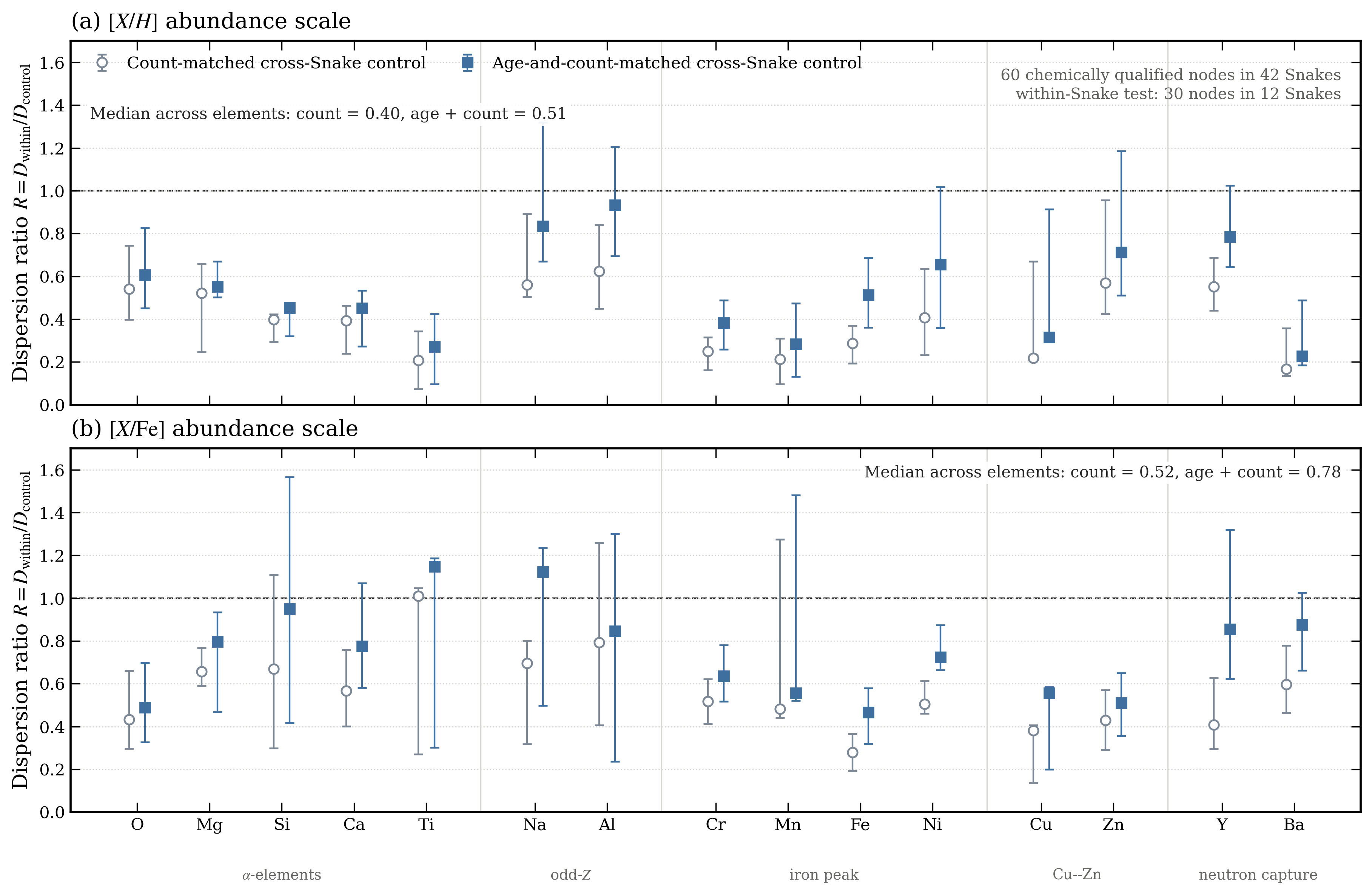}
    \caption{
    Chemical-coherence comparison of GALAH-sampled Snake base nodes against
    catalogue-wide control ensembles. Open circles and filled squares show,
    respectively, the ratio
    \(R=D_{\rm within}/D_{\rm control}\) between the within-Snake robust
    node-abundance dispersion and that of count-matched and
    age-and-count-matched cross-Snake controls. Values below unity indicate
    lower abundance dispersion among nodes assigned to the same Snake relative
    to the corresponding catalogue-wide control. Error bars show the bootstrap
    16th--84th percentile intervals of the ratios. Panel (a) presents the
    \([X/H]\) abundance scale, for which the median ratios across elements are
    0.40 and 0.51 for the count-matched and age-and-count-matched controls,
    respectively. Panel (b) presents the \([X/{\rm Fe}]\) scale, for which the
    corresponding median ratios are 0.52 and 0.78.
    The chemically qualified sample contains 60 nodes in 42 Snakes;
    the within-Snake dispersion test uses the 30 nodes belonging to the
    12 Snakes with at least two chemically qualified nodes. Elements are
    arranged into broad nucleosynthetic groups: \(\alpha\)-elements
    (O, Mg, Si, Ca, and Ti), odd-\(Z\) elements (Na and Al), iron-peak elements
    (Cr, Mn, Fe, and Ni), Cu--Zn, and neutron-capture elements (Y and Ba).
    All element-wise ratios draw on the same parent sample, with
    element-dependent availability, and are therefore statistically dependent. The median-across-elements
    values are descriptive aggregate summaries rather than combinations of
    fifteen independent tests.
    }
    \label{fig:snake_chemical_validation}
\end{figure*}

The fifteen element-wise ratios share a parent sample of
30 chemically sampled nodes in 12 Snakes, with element-dependent
availability, and are therefore not independent tests. Figure~\ref{fig:snake_chemical_validation}
groups the elements into broad nucleosynthetic families to display their
species-to-species behaviour. Elements such as Cu, Zn, Y, and Ba can carry
larger uncertainties or stronger evolutionary-stage systematics than many
of the \(\alpha\)- and iron-peak elements. We therefore use the
median-across-elements ratios only as descriptive aggregate summaries and
do not treat any individual element as an independent confirmation of
chemical coherence.

This chemical comparison is deliberately independent of the SRI
construction and is not used to define the reliability classes. The
catalogue-wide controls show a consistent directional contrast toward lower
abundance dispersion among same-Snake nodes, while the strict local
comparison is limited to only two independent matched blocks and is not
statistically decisive. We therefore treat chemistry as a sparse auxiliary
consistency diagnostic rather than as independent confirmation of a common
origin or as a catalogue-wide selection criterion.

\subsection{Re-evaluating the Original Stellar Snake: Kinematic Subdivision and Orbital Traceback}
\label{subsec:re-evaluating_snake}

Previous targeted studies of the original Stellar Snake identified
substantial internal structure: \citet{wang2022stellar} reported four
kinematic groups within the broad Part\,I--Part\,II configuration, and
\citet{yang2024stellar} further resolved the region using its stellar-population
structure. The homogeneous source-level search used here is constructed
independently of those earlier partitions and does not use their boundaries
as input. We therefore use this region as a consistency test of whether the
previously reported subdivision can be recovered independently within the
3.1~kpc catalogue.

In the final catalogue, the original projected region is represented by
three catalogue groups. Snake~0 corresponds to the original Part\,I,
Snake~1 traces the main body of the original Part\,II, and Snake~2 represents
the Trumpler~10-centred branch. The independently identified Snake~III
comparison region is represented by Snake~3.

\begin{figure*}[htpb]
    \centering
    \includegraphics[width=0.95\textwidth]{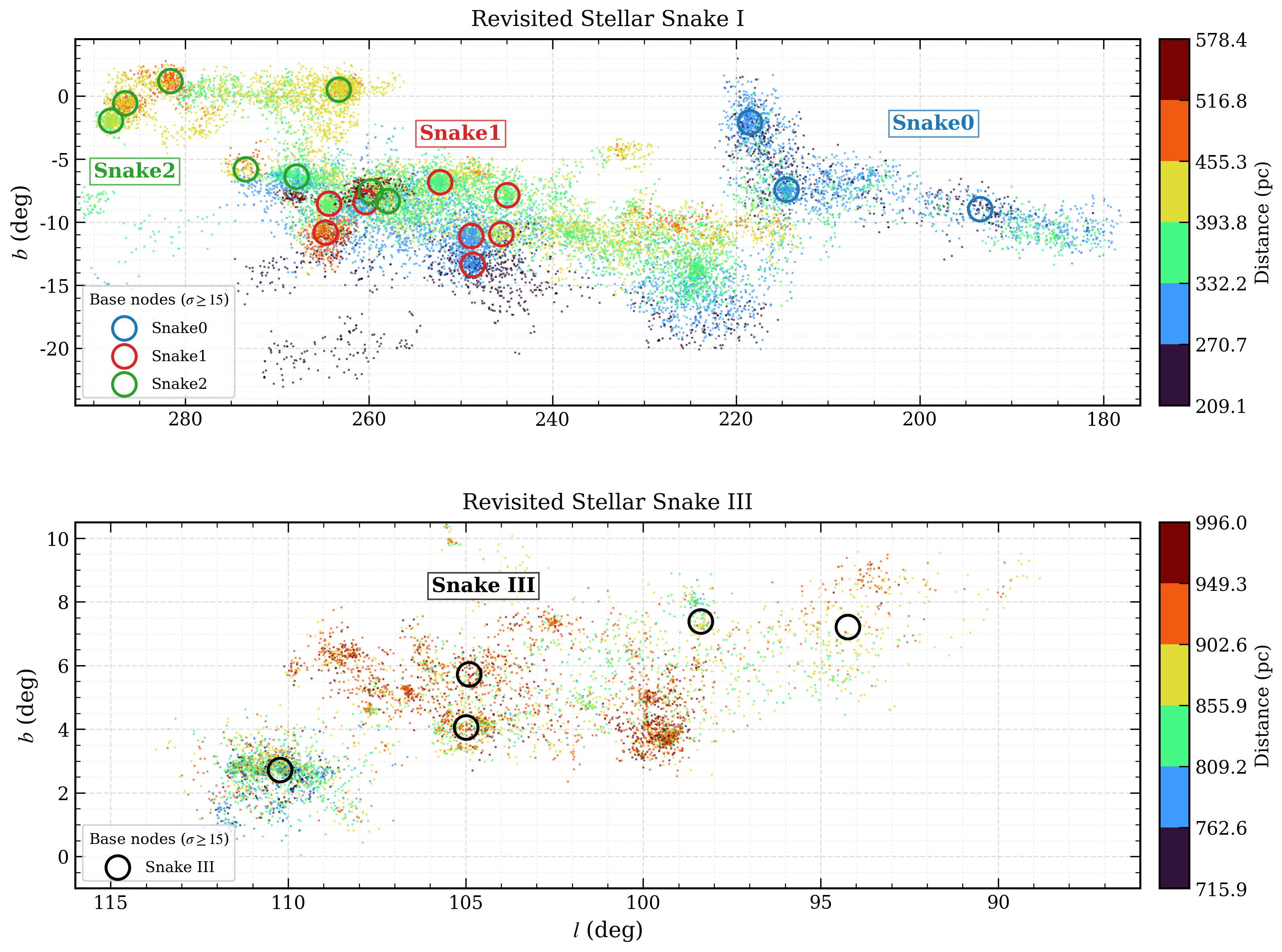}
    \caption{
Sky-projected distributions of the revisited Stellar Snake structures.
The upper panel shows the original Stellar Snake region from
\citet{wang2022stellar}, decomposed into Snake~0, Snake~1, and the
Trumpler~10-centred Snake~2. The lower panel shows the independently
identified Snake~III region \citep{li2026stellar} for methodological
comparison only. In both panels, member stars are colour-coded by
heliocentric distance with independent colour scales. Coloured open
circles denote the displayed high-significance base nodes with
\(\sigma>15\).
    }
    \label{fig:snake_revisit_lb}
\end{figure*}

\begin{figure}[htpb]
    \centering
    \includegraphics[width=0.45\textwidth]{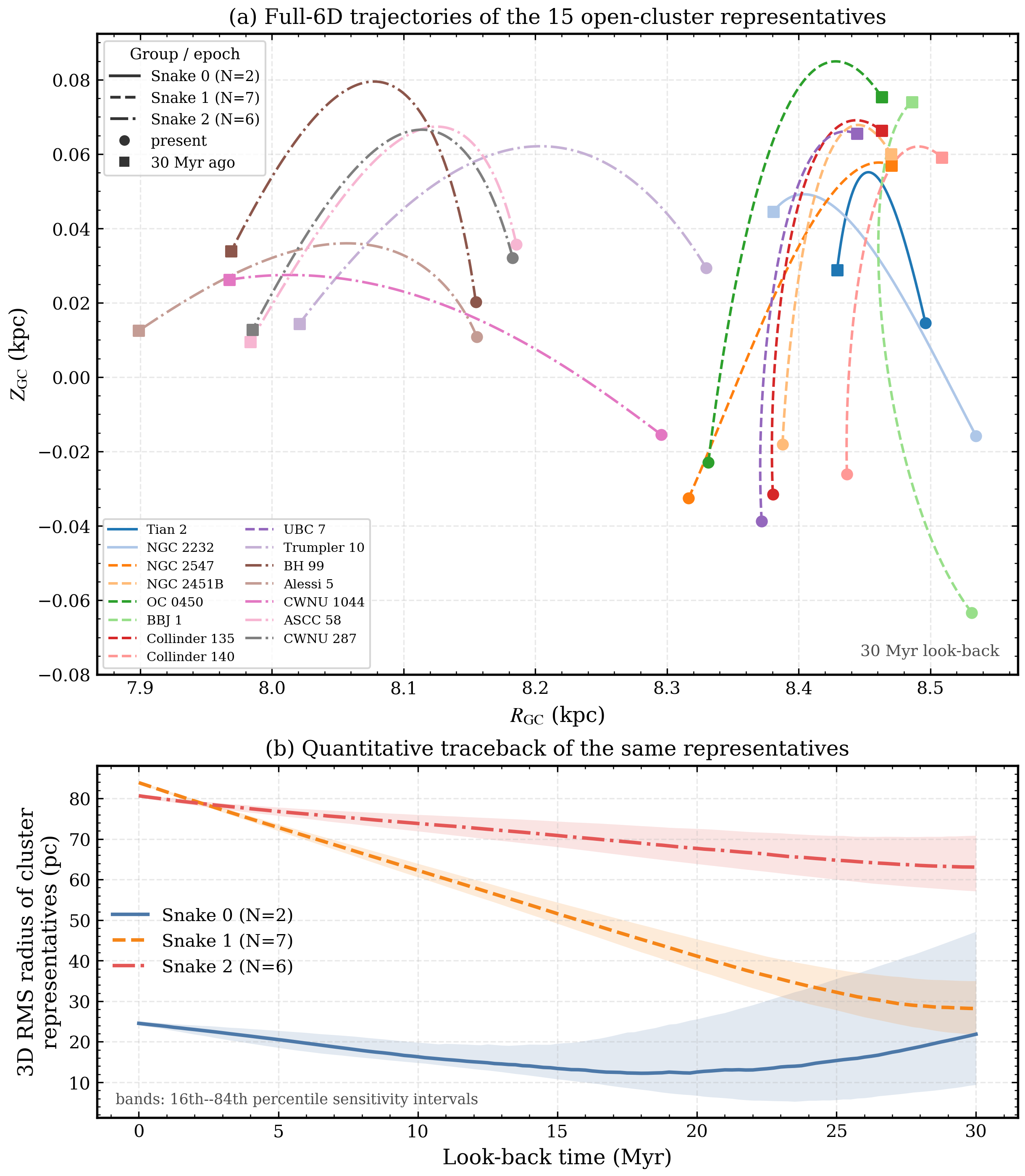}
    \caption{Full-6D orbital traceback of the 15 named Hunt--Reffert open-cluster
    representatives assigned to Snake~0, Snake~1, and Snake~2 over
    30 Myr. Panel (a) shows the trajectories projected onto the
    Galactocentric \(R\)--\(Z\) plane. Solid, dashed, and dash-dot lines
    denote representatives of Snake~0, Snake~1, and Snake~2, respectively;
    filled circles mark present-day positions, and squares mark the
    30 Myr look-back positions. Panel (b) shows the Monte Carlo median three-dimensional RMS radius of the cluster representatives about the instantaneous group centroid as a function of look-back time. The groups contain
    \(N_g=2\), 7, and 6 representatives, respectively. Shaded bands show
    the 16th--84th percentile sensitivity intervals obtained from 500
    member-bootstrap and quoted-error-perturbation realisations. Snake~1
    contracts substantially more strongly in traceback than Snake~2,
    although the minima of both curves occur at the 30 Myr integration
    boundary and are not interpreted as convergence ages.
    }
    \label{fig:orbits}
\end{figure}

Figure~\ref{fig:snake_revisit_lb} shows the present-catalogue sky projection. In the
upper panel, the original Stellar Snake region is resolved into Snake~0,
Snake~1, and the Trumpler~10-centred Snake~2. Member stars are colour-coded
by heliocentric distance, and open circles mark the displayed
high-significance base-node subset with \(\sigma > 15\). The quantitative dynamical comparison is presented in
Figure~\ref{fig:orbits} using full-6D orbital traceback.

The lower panel shows the independent Stellar Snake~III region
\citep{li2026stellar} using the same plotting scheme. It is included as a
methodological comparison only. It is not treated as a physical extension of
the original \citet{wang2022stellar} structure, nor as a member of Snake~0,
Snake~1, or Snake~2.

\begin{itemize}
    \item \textbf{Snake~0 (formerly Part\,I):} associated with NGC~2232 and Tian~2.
    \item \textbf{Snake~1 (main body of the former Part\,II):} associated with NGC~2547, NGC~2451B, OC~0450, BBJ~1, Collinder~135, Collinder~140, and UBC~7.
    \item \textbf{Snake~2 (Trumpler~10-centred branch):} associated with Trumpler~10, BH~99, Alessi~5, CWNU~1044, ASCC~58, and CWNU~287.
\end{itemize}

We further compare these three groups with the Hunt--Reffert open-cluster
catalogue at the Snake-group level. The resulting match fractions are listed
in Table~\ref{tab:cross_match}.

\begin{table}[htpb]
\centering
\caption{
Snake-group-level match fractions with the \citet{hunt2023improving}
open-cluster catalogue. For each Hunt--Reffert cluster, the
``Cross-matched'' column gives the number of catalogued members recovered in
the corresponding Snake group, divided by the total catalogued membership.
}
\label{tab:cross_match}
\scalebox{0.65}{
\begin{tabular}{llccc}
\hline \hline
\textbf{Group} & \textbf{Cluster Name} & \textbf{Hunt ID} &
\textbf{Cross-matched} & \textbf{Ratio (\%)} \\
\hline
\multirow{2}{*}{Snake~0}
& Tian 2 (ZHBJZ 1) & 7156 & 472 / 490 & 96.3 \\
& NGC 2232 & 4524 & 275 / 286 & 96.2 \\
\hline
\multirow{7}{*}{Snake~1}
& NGC 2547 & 4590 & 628 / 675 & 93.0 \\
& NGC 2451B & 4873 & 589 / 615 & 95.8 \\
& OC 0450 & 5015 & 389 / 398 & 97.7 \\
& BBJ 1 (LP 2383) & 1745 & 387 / 422 & 91.7 \\
& Collinder 135 & 1323 & 207 / 209 & 99.0 \\
& Collinder 140 & 1324 & 195 / 203 & 96.1 \\
& UBC 7 (Alessi 36) & 86 & 194 / 198 & 98.0 \\
\hline
\multirow{6}{*}{Snake~2}
& Trumpler 10 & 6048 & 1152 / 1425 & 80.8 \\
& BH 99 & 150 & 577 / 643 & 89.7 \\
& Alessi 5 & 70 & 490 / 571 & 85.8 \\
& CWNU 1044 & 576 & 327 / 377 & 86.7 \\
& ASCC 58 & 25 & 206 / 236 & 87.3 \\
& CWNU 287 & 456 & 34 / 36 & 94.4 \\
\hline
\end{tabular}
}
\end{table}

To quantify the past spatial extent of the open-cluster
component, we integrate the 15 Hunt--Reffert representatives
listed in Table~\ref{tab:cross_match} backward over 30 Myr.
All are classified as \texttt{Type=o}, with recovered membership
fractions of 80.8\%--99.0\%. The traceback characterises this
reference sample rather than every constituent base node. Each representative is defined from the intersection
between the corresponding Hunt--Reffert cluster membership list and the full
Snake group listed in Table~\ref{tab:cross_match}. Each trajectory is
initialized from the median 6D phase-space coordinates of the matched stars:
median sky position, median inverse-parallax distance
\(\mathrm{median}(1/\varpi)\), median proper motion, and median radial
velocity. The integrations follow the same dynamical convention described in
Section~\ref{sec:data}.
At each look-back time, we quantify the spatial scale of group \(g\)
using the three-dimensional RMS radius of its cluster representatives about
their instantaneous group centroid,
\begin{equation}
R_{{\rm rms},g}(t)=
\left[
\frac{1}{N_g}
\sum_{j=1}^{N_g}
\left\|
\boldsymbol{x}_j(t)-\overline{\boldsymbol{x}}_g(t)
\right\|^2
\right]^{1/2},
\end{equation}
where \(\boldsymbol{x}_j(t)\) is the Galactocentric Cartesian position of
representative \(j\). The 16th--84th percentile sensitivity intervals are
obtained from 500 member-bootstrap and quoted-error-perturbation
realisations. These intervals describe sensitivity to member sampling and
independently perturbed quoted measurement errors; Gaia error
covariances and uncertainties in the adopted Galactic potential
are not included.

Figure~\ref{fig:orbits} shows both the Galactocentric \(R\)--\(Z\)
trajectories and the three-dimensional RMS-radius evolution. The Monte Carlo median RMS radius of Snake~0,
which contains only two representatives, changes from
\(24.5\,[24.1,25.2]\) pc at the present day to
\(21.9\,[9.4,47.2]\) pc at 30 Myr; its broad traceback interval does not
support a robust convergence inference. Snake~1 decreases from
\(83.9\,[83.6,84.2]\) pc to \(28.2\,[21.8,35.0]\) pc, whereas
Snake~2 decreases from \(80.7\,[80.2,81.0]\) pc to
\(63.1\,[57.1,70.9]\) pc. Thus, despite their similar present-day RMS
scales, Snake~1 and Snake~2 show quantitatively different full-6D traceback
behaviour.

This behaviour is consistent with the independent interpretation of
\citet{swiggum2024most}, who argued that the M6 family, including
Trumpler~10, and the Cr~135 family were spatially separated during formation
and began to overlap only during the past \(\sim15\) Myr. Our catalogue-level
reinterpretation is therefore not that the original Stellar Snake was a
single monolithic birth structure, but that the same projected region
contains multiple young kinematic families. The full-6D traceback and RMS-radius evolution quantitatively support
the subdivision into components with different traceback histories, but do
not by themselves establish distinct birth environments. Because the
minimum RMS scales of Snake~1 and Snake~2 occur at the 30 Myr integration
boundary, we do not infer a convergence age, birth radius, or common
birthplace.

\paragraph{Catalogue-wide comparison with the Swiggum families.}
To test whether the agreement extends beyond the named representatives, we
cross-matched the final catalogue with the three principal young-cluster
families of \citet{swiggum2024most}. These family assignments were derived
from full six-dimensional motions and orbital traceback and were not used
in our catalogue construction. Although the studies share Gaia measurements, the external family labels
provide a separate check on our linking boundaries. Of the 155 published family members, 138 have usable source-level membership lists. After applying the common source-level cuts of \(d_\varpi\leq3.1\)~kpc and \(\varpi/\sigma_\varpi>10\),
128 clusters retain at least 30 members. An external cluster is defined as mapped when at least 30\% of its retained members are assigned to the same final Snake.

Under this definition, 26 of 38 comparable Cr~135-family clusters map
to five Snakes, 19 of 31 M6-family clusters map to seven Snakes, and 41 of
59 \(\alpha\)~Per-family clusters map to thirteen Snakes. In total, 86 of
128 comparable clusters map to 25 final Snakes, and none of these Snakes
contains mapped clusters from more than one of the three principal families.
This catalogue-wide comparison extends the Snake~1/Snake~2 boundary agreement
beyond the named anchor clusters and reveals no detected cross-family
over-merging in the comparable sample. The numbers of Snakes associated with
each family are reported as partition multiplicities rather than as a formal
fragmentation rate, because the Swiggum families and Stellar Snakes are not
defined at the same hierarchy level.

\section{Summary}
\label{sec:summary}

We have presented a Gaia-source-level catalogue of Stellar Snake
complexes in a nominal 3~kpc volume, using a 3.1~kpc selection boundary.
The pipeline extracts statistically significant, non-overlapping FoF
nodes, estimates their parameters with a PointNet regressor, and links
them using positions, kinematics, ages, and derived dynamical quantities.
Its principal products are the source-level-to-complex catalogue,
homogeneous node-parameter estimates, and a continuous SRI measure of
spatial--kinematic compatibility. A base node may be a compact cluster
or a loose association; a Snake is an association of two or more such
overdensities, not necessarily a single bound or co-natal object.

The final catalogue contains 1,180 Stellar Snake candidates,
with 4,196 base nodes and 692,408 unique Gaia member sources,
including bridge stars. External comparisons with open-cluster catalogues and spectroscopic
metallicity benchmarks show broad agreement for age and extinction.
The distance-modulus comparison reflects shared parallax information,
while metallicities show limited external rank-order consistency. The continuous full SRI, reported with its A/B/C evidence group,
is summarised by empirical recommended grades: 727 Gold, 365 Silver,
and 88 Bronze. Users may select other thresholds for their purposes.
KPD separately flags 79 of 636 assessed systems; 544 are unassessed,
and core scores do not replace full-score grades.
The Palma pairs show \({\rm AUC}=0.950\), with an optical-control
false-positive fraction of 1.72\% at the recommended Gold boundary.
These are nearby-pair consistency statistics, not a probability
calibration across all evidence groups.

Motivated by concerns raised for extended stellar strings,
we examined auxiliary kinematic and chemical diagnostics. At the kinematic level, larger \(f_{\rm cl}\) is associated with
smaller observed member-level RV scatter, including after controls
for distance and median \(G\). High-\(f_{\rm cl}\) nodes approach
the observed RV scale of Hunt--Reffert open clusters.
These results support an empirical ranking of member-sample cleanliness,
without calibrating absolute membership purity or determining
intrinsic velocity dispersions. At the chemical level,
chemically sampled base nodes in the same Stellar Snake are more similar
across fifteen GALAH abundance elements than count-matched and
age-count-matched cross-Snake controls, while the descriptive node-pair doppelganger frequencies are 14.3\%
on the \([X/H]\) scale and 17.1\% on the \([X/\mathrm{Fe}]\) scale. The strict local comparison is limited to two matched blocks and is
not statistically decisive. Chemistry therefore remains an auxiliary
consistency diagnostic rather than independent evidence for a common
chemical origin.

The catalogue reveals several population-level trends, including a
descriptive age--metallicity pattern in the jointly inferred
parameters and a declining upper envelope of observed member-star
entries toward older inferred ages. \add{The matched-cluster comparison shows no significant excess
proximity to spiral-arm centrelines. The subset selected near the
projected Radcliffe Wave shows a vertical morphological correspondence
(Spearman \(\rho=0.665\)).} Re-examining the original Stellar Snake
recovers its previously reported broad subdivision into multiple
kinematic components, while
the recovery of the spatially independent Stellar Snake III region illustrates
the transferability of the same source-level-to-complex framework.

The catalogue remains selection limited: its grouping resolution
depends on the clustering settings, the SRI evidence varies with
entity multiplicity and RV support, and neither completeness nor
cross-group physical purity has an end-to-end calibration.
Incomplete RV coverage, sparse spectroscopy, unresolved binaries,
extinction, and the reduced detectability of dispersed populations
limit the physical interpretation. Representative associated and
unrelated simulations embedded in the observed field, propagated
through the full detection and assembly pipeline, are needed to
calibrate recovery, fragmentation, and score-based selections.
Future astrometry and spectroscopy can strengthen these tests.

\section*{Data and Code Availability}
The machine-readable catalogue schema is given in
Tables~\ref{table:snake_complex}--\ref{table:snake_member}.
The machine-readable catalogues and accompanying README
are supplied with the revised submission.
\add{The hierarchical-output and metric-weight modifications to
\textsc{Rockstar} are documented in
\citet[Chapter~2, Appendix~2.6]{tian2017hide}.
The SRI reference implementation is available from
the code repository\footnote{\url{https://github.com/AstroHDU/Snake/}}.}

{\it Acknowledgements.}
The authors thank Feng Wang and Dolev Bashi for the helpful discussions. H.J.T. thanks the support from the NSFC grant (No. 12373033) and the Key Project of Zhejiang Provincial Natural Science Foundation (No. ZCLZ25A0301). 

%

\vspace{5mm}





\appendix
\renewcommand*{\theHequation}{app.\Alph{section}.\arabic{equation}}

\section{Quantitative Definition of the SRI}
\label{app:sri}

The SRI combines spatial and velocity member continuity
(Mem), relative longest-edge centre discontinuities (Cen), and,
where RV support permits, a global spatial--velocity relation (Cross).
KPD is computed separately and does not enter the score.
The following equations describe the configuration used for the
published scores; all positions, velocities, and required uncertainties
are reconstructed from the source table.

\subsection{Entity Representation and RV Eligibility}
\label{app:sri_contraction}
\label{app:sri_rv}

The FoF hierarchy can divide one extended overdensity into several
adjacent base nodes. Before scoring, locally compatible fragments are
therefore allowed to contract into common graph entities. For base nodes
\(a\) and \(b\), let \(g_{ab}\) be their minimum member-to-member Cartesian
separation and let \(\ell_a\) and \(\ell_b\) be the median internal
XYZ-MST edge lengths of the two nodes. Their spatial-contact ratio is
\begin{equation}
    R_X(a,b)
    =
    \frac{g_{ab}}{\ell_a+\ell_b}.
    \label{eq:sri_entity_rx}
\end{equation}

Velocity compatibility is evaluated in a common tangent-velocity
plane. For projection direction
\(\hat{\mathbf u}_\theta=(\cos\theta,\sin\theta)\), let
\(m_a(\theta)\) and \(s_a(\theta)\) denote the projected median and MAD
of node \(a\). We define
\begin{equation}
    R_V(a,b)
    =
    \max_\theta
    \frac{|m_a(\theta)-m_b(\theta)|}
    {s_a(\theta)+s_b(\theta)} .
    \label{eq:sri_entity_rv}
\end{equation}

A local pair is compatible when \(R_X\le1\) and \(R_V\le1\).
For multi-node entities, spatial compatibility uses the minimum
cross-group \(R_X\), whereas velocity compatibility uses the maximum
cross-group \(R_V\). This allows spatially connected fragments to merge
while preventing velocity-space chaining. The angular maximisation is
evaluated on 720 directions; this is a numerical resolution rather than
a physical threshold. If natural contraction collapses the entire Snake
to one entity, the original base-node representation is retained for
scoring.

When both nodes have at least three valid RV measurements,
an otherwise compatible merger is vetoed if their velocity difference
projected along the common radial direction exceeds three times its
combined systemic uncertainty. Missing RV information does not itself
veto a merger.

Source positions use \(d=1000/\varpi\) pc for \(\varpi\) in mas.
The local tangential velocities are
\(V_k=\kappa\mu_k/\varpi\), with
\(\kappa=4.74047\) in the adopted units and
\(k\in\{\alpha *,\delta\}\). Their first-order variances are
\begin{equation}
\epsilon_{V_k}^{2}=\kappa^2\left[
\frac{\sigma_{\mu_k}^{2}}{\varpi^2}
+\frac{\mu_k^2\sigma_{\varpi}^{2}}{\varpi^4}
-\frac{2\mu_k\rho_{\varpi,\mu_k}
\sigma_{\varpi}\sigma_{\mu_k}}{\varpi^3}\right].
\label{eq:sri_source_tangent_error}
\end{equation}
The release retains both parallax--proper-motion correlations.
The current component-error approximation does not propagate the full
five-parameter astrometric covariance, sky-position errors, or
cross-component covariance into the systemic-error model.
The spatial entity centre is the node-member-count-weighted mean of
the constituent nodes' componentwise median XYZ positions; it is not
in general the median position of the pooled members.

After contraction, each entity's three-dimensional velocity centre
is estimated separately from the member-cloud velocity transport.
The medians of the valid source-level
\(V_{\alpha *}\), \(V_\delta\), and RV measurements are rotated into
\((U,V,W)\) at the entity's mean sightline. Each component uses its own
available measurements; missing stellar RVs are not filled in.
For component \(k\), the adopted systemic uncertainty is
\begin{equation}
\begin{aligned}
\sigma_{e,k}^{2}
&=\left(\sum_{s\in e,k}\epsilon_{s,k}^{-2}\right)^{-1}
+\frac{s_{{\rm int},e,k}^{2}}{N_{e,k}},\\
\mathbf C_e&=\mathbf B_e
{\rm diag}(\sigma_{e,k}^{2})\mathbf B_e^{\mathsf T}.
\end{aligned}
\end{equation}
Here \(N_{e,k}\) is the number of valid measurements,
\(s_{{\rm int},e,k}^2=\max[0,s_{{\rm obs},e,k}^2-
{\rm med}(\epsilon_{s,k}^2)]\), with
\(s_{{\rm obs},e,k}=1.4826\,{\rm MAD}(V_{s,k})\).
\(\mathbf B_e\) rotates the local velocity basis into heliocentric
Galactic Cartesian coordinates, without a solar-motion correction.
This independent-component approximation describes uncertainty in the
representative centre, rather than the full member velocity dispersion
or an exact sampling covariance of the medians.

No missing RV is filled to construct a three-dimensional centre.
An entity enters the three-dimensional scores only when it has a finite
representative centre and at least three distinct sources with finite
RV measurements and positive, finite RV uncertainties,
\begin{equation}
    N_{{\rm RV},e}\ge3 .
    \label{eq:sri_rv3}
\end{equation}
Entities failing this condition remain in all member-cloud
calculations; they are excluded only from the three-dimensional centre
terms. Let \(M_{\rm RV}\) denote the number of RV-qualified entities.
We define evidence groups A, B, and C by
\(M_{\rm RV}<3\), \(M_{\rm RV}=3\), and \(M_{\rm RV}\ge4\),
respectively.

\subsection{Member and Centre Scores}
\label{app:sri_statistics}
\label{app:sri_channels}

For each member cloud, a robust scatter matrix \(\mathbf S\)
is estimated from pairwise differences using median-length clipping
and a pilot-whitening refinement. This describes member-cloud extent,
not centre uncertainty. Both clouds in a pair are transformed with
\(\mathbf H_{AB}^{-1}\), where
\(\mathbf H_{AB}=\mathbf S_A^{1/2}+\mathbf S_B^{1/2}\).
For dimension \(p\), their symmetric nearest-neighbour distance is
\begin{equation}
\begin{aligned}
d(A,B)
&=\frac{1}{2\sqrt{p}}\Biggl[
{\rm med}_{a\in A}\min_{b\in B}
\|\widetilde{\mathbf a}-\widetilde{\mathbf b}\|\\
&\qquad\quad+
{\rm med}_{b\in B}\min_{a\in A}
\|\widetilde{\mathbf b}-\widetilde{\mathbf a}\|
\Biggr].
\end{aligned}
\label{eq:sri_member_distance}
\end{equation}

The spatial and tangential-velocity responses are
\begin{equation}
\begin{aligned}
s_X(A,B)&=\left[1+\left(\frac{d_X(A,B)}{a_X}\right)^2\right]^{-1/2},
\quad a_X=2,\\
s_V(A,B)&=\exp[-d_V(A,B)].
\end{aligned}
\label{eq:sri_member_responses}
\end{equation}
The node-member scores \(S_{\rm mem,0}\) and \(V_{\rm mem}\)
are mean responses over their respective member-distance MSTs.
Velocity clouds use the common tangent-plane convention described
above; they do not define the three-dimensional scoring centres.

For companion support, each bridge source is assigned to the
nearest eligible entity member cloud within the same \texttt{id\_part}.
Let \(g_{\rm obs}\) be the spatial-score gain, evaluated on the fixed
node-member MST after adding these assignments. The rank \(q\) of this
gain is computed with half weight for ties among 100 controls rotating
each part's companion configuration about its heliocentric line of sight.
The adopted spatial score is
\begin{equation}
S_{\rm mem}=\operatorname{clip}_{[0,1]}
\left[S_{\rm mem,0}+\max(g_{\rm obs},0)\max(2q-1,0)\right].
\label{eq:sri_bridge_support}
\end{equation}
Without bridge sources, \(S_{\rm mem}=S_{\rm mem,0}\).
The correction cannot reduce node-only continuity. Bridge sources
enter neither the intrinsic entity widths used by Scen nor the velocity
centres. These line-of-sight rotations preserve source distances and
configuration geometry, but not the full angular extinction or density
field. Their gain rank measures additional spatial support,
not the probability of physical association.

Each Cen term tests only the longest edge of its physical
centre MST, relative to the remaining tree, rather than maximising
an anomaly over locally normalised edges. For the spatial tree,
let \(L_X\) be its longest edge, \(b_X\) the linearly interpolated
75th percentile of all other edge lengths, and
\(\hat{\mathbf u}\) the longest-edge direction. Define
\begin{equation}
\begin{aligned}
w_e&=(\hat{\mathbf u}^{\mathsf T}\mathbf S_e\hat{\mathbf u})^{1/2},\\
W_{\rm test}&=w_i+w_j,\\
W_{\rm ref}&={\rm med}_{(a,b)\ne(i,j)}(w_a+w_b),\\
B_X&=\max(b_X,W_{\rm test},W_{\rm ref}).
\end{aligned}
\label{eq:sri_spatial_cen_floor}
\end{equation}
\begin{equation}
S_{\rm cen}=\sqrt{\min(1,B_X/L_X)}.
\label{eq:sri_spatial_cen}
\end{equation}
All widths are projected along the same tested direction and
use node members only. Coincident centres, \(L_X=0\), give
\(S_{\rm cen}=1\).

For the velocity tree on the RV-qualified heliocentric
\((U,V,W)\) centres, let \(L_V\) be the longest edge and \(b_V\)
the 75th percentile of the remaining positive edge lengths.
Set \(\delta=L_V-b_V\). With the nominal tree and quantile ordering
held fixed, propagate the shared-endpoint gradients through the
entity centre covariances:
\begin{equation}
\begin{aligned}
\sigma_\delta^2&=
\sum_e(\nabla_{\mathbf v_e}\delta)^{\mathsf T}
\mathbf C_e(\nabla_{\mathbf v_e}\delta),\\
d&=\frac{\max(\delta,0)}{b_V},
\qquad q_\delta=\frac{\sigma_\delta^2}{b_V^2}.
\end{aligned}
\label{eq:sri_cen_excess}
\end{equation}
\begin{equation}
\begin{aligned}
d_{\rm eff}&=
\left[\min(d^2,1)+\frac{\max(d^2-1,0)}{1+q_\delta}\right]^{1/2},\\
V_{\rm cen}&=(1+d_{\rm eff})^{-1/2}.
\end{aligned}
\label{eq:sri_cen_scores}
\end{equation}
The quantile interpolation and shared endpoints enter the
propagation; the uncertainty does not rebuild the MST.
Zero uncertainty recovers \(V_{\rm cen}=\sqrt{b_V/L_V}\).
For \(d>1\), even arbitrarily large uncertainty cannot raise the
score above \(1/\sqrt{2}\); only excess discontinuity evidence is
attenuated. Identical velocity centres give unity. A positive longest
edge without a positive reference scale is explicitly invalid rather
than assigned an arbitrary scale. Both Cen scores enter the aggregate
only for \(M_{\rm RV}\ge3\): Scen then uses all spatial entities,
whereas Vcen uses the RV-qualified subset.

\subsection{Cross Relation and SRI Aggregation}
\label{app:sri_cross}
\label{app:sri_aggregation}

For Group-C systems, Cross measures whether the entity velocity
field is compatible with a single first-order spatial relation.
Positions are centred and RMS-scaled. For each held-out entity \(i\),
an affine model
\begin{equation}
    \widehat{\mathbf v}(\mathbf z)=\mathbf b+\mathbf G\mathbf z
\end{equation}
is fitted to the remaining entities by minimising
\begin{equation}
    \sum_{j\ne i}\omega_j^{(-i)}
    \|\mathbf v_j-\mathbf b-\mathbf G\mathbf z_j\|^2
    +\alpha N_{\rm train}\|\mathbf G\|_F^2,
    \label{eq:sri_affine}
\end{equation}
with \(\alpha=0.10\) and an unpenalised intercept.
Influence weights are determined within the training set and moderated
to retain adequate effective support; the held-out velocity is not
used to determine these weights.

Let \(\mathbf Q_{ij}=\delta_{ij}\mathbf I-
\mathbf W_{ij}^{(-i)}\) be the residual operator with the training
influence weights held fixed, and let
\(A_i=1+\sum_{j\ne i}(w_{ij}^{(0)})^2\) be the amplification
from the corresponding unweighted scalar ridge design. Define
\begin{equation}
\mathbf r_i=\frac{\sum_j\mathbf Q_{ij}\mathbf v_j}{\sqrt{A_i}},
\qquad\mathbf E_i=\frac{\sum_j\mathbf Q_{ij}\mathbf C_j
\mathbf Q_{ij}^{\mathsf T}}{A_i}.
\label{eq:sri_cross_residual}
\end{equation}
The covariance propagation is conditional on the fitted weights
and nominal positions. In the target entity's sightline frame,
write the radial residual and variance as \(r_r,\sigma_r^2\),
and diagonalise the covariance within the actual two-dimensional
tangent plane to obtain \(r_{t,k},\sigma_{t,k}^2\).
With \(\tau^2=h_{\rm cross}^2/3\), define
\begin{align}
\phi(r,\sigma^2)&=\min(r^2,\tau^2)+
\frac{\tau^2}{\tau^2+\sigma^2}\max(r^2-\tau^2,0),\\
\widetilde\ell_r&=\min(r_r^2,\tau^2)+\notag\\
&\qquad
\frac{\tau^2}{\tau^2+\sigma_r^2}
\sqrt{\frac{r_r^2}{r_r^2+\sigma_r^2}}
\max(r_r^2-\tau^2,0),\\
g_i&=1-\frac{M_{\rm RV}-4}{M_{\rm RV}}
\left[1-\left(1+\frac{\|\mathbf r_t\|^2}{2\tau^2}\right)^{-1/2}\right],\\
\ell_i&=\sum_{k=1}^{2}\phi(r_{t,k},\sigma_{t,k}^2)
+r_r^2-g_i(r_r^2-\widetilde\ell_r).
\end{align}
Zero-over-zero residual fractions use their continuous limiting
loss. Measurement uncertainty attenuates only the excess above
\(\tau^2\); radial attenuation also depends on tangential compatibility
and redundancy beyond four entities. The aggregate residual scale is
\begin{equation}
D_{\rm cross}=\frac{1}{M_{\rm RV}}\sum_i\sqrt{\ell_i}.
\label{eq:sri_dcross}
\end{equation}
We map this velocity scale to
\begin{equation}
    C_{\rm cross}=\left[
    1+\left(\frac{D_{\rm cross}}{h_{\rm cross}}\right)^2
    \right]^{-1/2},
    \qquad h_{\rm cross}=5.0~{\rm km\,s^{-1}}.
    \label{eq:sri_cross_quality}
\end{equation}
Thus, Cross averages adjusted residual magnitudes without
capping the largest loss. Four RV-qualified entities permit the
regularised calculation, but do not guarantee that the spatial
prediction is identifiable without regularisation.
The resulting score measures compatibility under the adopted
uncertainty model; weak constraints alone are not evidence of
a physically reliable association.

The spatial and velocity pillars are
\begin{equation}
\begin{aligned}
S&=\begin{cases}
S_{\rm mem},&M_{\rm RV}<3,\\
\sqrt{S_{\rm mem}S_{\rm cen}},&M_{\rm RV}\ge3,
\end{cases}\\
V&=\begin{cases}
V_{\rm mem},&M_{\rm RV}<3,\\
\sqrt{V_{\rm mem}V_{\rm cen}},&M_{\rm RV}\ge3.
\end{cases}
\end{aligned}
\end{equation}
\begin{equation}
R_0=\frac{2SV}{S+V},\qquad
{\rm SRI}=\begin{cases}
R_0,&M_{\rm RV}<4,\\
R_0C_{\rm cross},&M_{\rm RV}\ge4.
\end{cases}
\label{eq:sri_final}
\end{equation}
Zero-score boundaries use the continuous limit. The 0.70 and 0.50
grade boundaries are empirical recommendations on this continuous
scale (Section~\ref{sec:snake_reliability}), not calibrated
probability thresholds.

\subsection{Separate Partition Diagnostic and Core Score}
\label{app:sri_outlier}

KPD is a separate geometric partition diagnostic on the same
scoring entities, without an SRI-gain requirement or direct uncertainty
attenuation of its edge ratios. It first partitions all spatially valid
entities using the spatial member-cloud distance matrix. Within each
spatial component, an RV-free tangential test can separate a side
consisting entirely of RV-limited entities; it does not repartition
the RV-qualified set. The qualified entities are then partitioned by
their physical three-dimensional velocity-centre distances.

\add{We adopt \(P>3\) as an empirical criterion, guided by
manual assessment of spatial and kinematic partitioning
results.}
For a component with at least three entities, the longest MST
edge is removed only when
\begin{equation}
\begin{aligned}
P&=\frac{L_{\max}}{L_{\rm ref}}>3,\\
L_{\rm ref}&=\begin{cases}
\max(L_{\rm second},1/\sqrt{p}),&\shortstack[l]{member-cloud\\distances},\\
L_{\rm second},&\shortstack[l]{three-dimensional\\velocity centres}.
\end{cases}
\end{aligned}
\label{eq:sri_outlier_prominence}
\end{equation}
Here \(p=3\) for spatial clouds and \(p=2\) for RV-free
velocity clouds. The floor is the unit radius in the same shared
\(\mathbf H\) metric after the \(1/\sqrt p\) normalisation, not an
uncertainty interval or the median internal pair separation.
Both tested- and reference-edge width floors have this value in the
current prescription. The procedure removes one edge at each accepted
step and recurses; components smaller than three entities are not
further tested.

After the three-dimensional split, RV-limited entities not
already separated by the tangential test are assigned to the closest
component using RV-free cloud distances to each component's fixed
RV-qualified member cloud. This does not invent RV measurements.
Exact distance ties remain unresolved. KPD is 1 if any accepted split
occurs and 0 if an assessment produces no split; a system with fewer
than three scoring entities has no KPD assessment.

For KPD=1, the unique largest component is selected by entity
count, with node-member count breaking ties; bridge sources do not
enter the tie breaker. If it contains at least two entities and all
assignments are unambiguous, its full calculation is repeated using
only retained-node sources and compatible bridge candidates, without
another entity contraction. Excluded node members are not relabelled
as bridge stars. The core score, evidence group, and status are
reported separately. A changed evidence group prevents interpreting
core-minus-full SRI as a like-for-like calibrated reliability gain.
The RV-support limit, merge veto, KPD threshold, ridge strength,
response scales, and recommended grade boundaries are explicit
operational conventions, not statistical confidence levels.

\section{Residual Correlations among PointNet Outputs}
\label{app:pointnet_residuals}

We evaluated the final PointNet model on 20,000 synthetic
stellar populations not used to train its weights, with age
and metallicity sampled independently. For each system,
\(q_{\rm pred}\) is the mean of 100 paired MC-dropout and
uniform member-resampling predictions, with Batch
Normalization statistics fixed. Correlations are computed
across these systems using the residuals
\begin{equation}
    \Delta q=q_{\rm pred}-q_{\rm true}.
\end{equation}

The true and predicted age--metallicity correlations are
both close to zero: \(R_{\rm true}=0.0118\) and
\(R_{\rm pred}=0.0022\), with corresponding Spearman
coefficients of \(S_{\rm true}=0.0118\) and
\(S_{\rm pred}=0.0019\). Their residuals are positively
correlated, with
\begin{equation}
    R(\Delta\log t,\Delta[{\rm Fe/H}])=0.3391.
\end{equation}
The Fisher 95\% confidence interval is \([0.3267,0.3513]\),
and the corresponding Spearman coefficient is \(S=0.2543\).
A two-sided test using 10,000 permutations with the \(+1\) correction
gives \(p<10^{-4}\).

\begin{figure*}[htbp]
    \centering
    \includegraphics[width=0.92\textwidth]
    {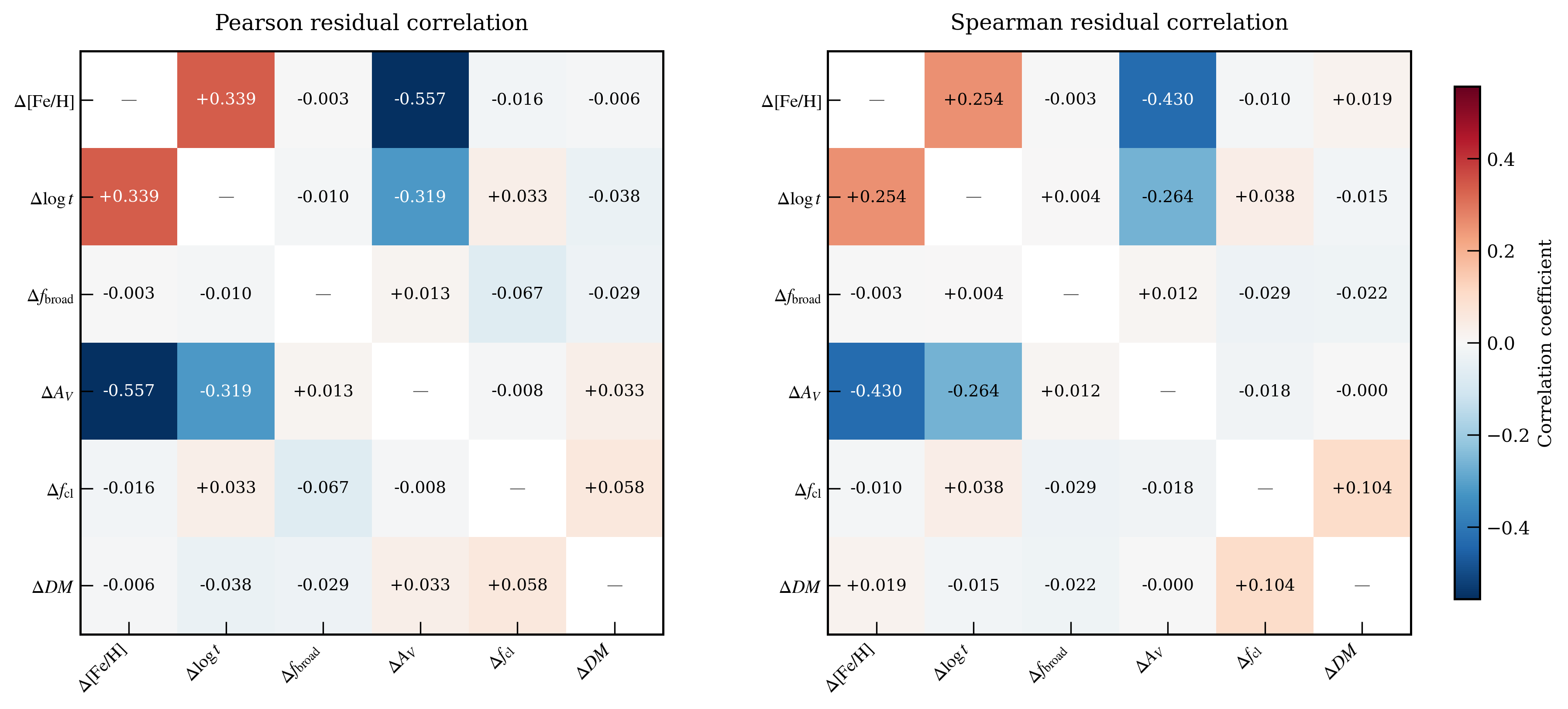}
    \caption{Pearson (left) and Spearman (right) residual correlations
    among the six PointNet outputs for 20,000 synthetic stellar
    populations not used to train the final model's weights. Residuals are predicted minus true values,
    using the mean of 100 paired MC-dropout and uniform member-resampling
    predictions per system, with Batch Normalization statistics fixed.}
    \label{fig:pointnet_residual_correlation}
\end{figure*}

Figure~\ref{fig:pointnet_residual_correlation} presents the complete
six-output residual-correlation matrices. Correlations with extinction
include \(R(\Delta[{\rm Fe/H}],\Delta A_V)=-0.557\) and
\(R(\Delta\log t,\Delta A_V)=-0.319\).
These results demonstrate that the prediction errors are not independent;
they do not establish that the catalogue-level age--metallicity pattern
is unbiased.

\bibliography{SnakeV}{}
\bibliographystyle{aasjournal}


\begin{table*}[htbp]
\nolinenumbers
\centering
\caption{Columns of the Stellar Snake Complex Catalogue}
\label{table:snake_complex}
\small
\begin{tabular}{c l l l}
\toprule\toprule
 & Column & Unit & Description\tabularnewline
\midrule
1 & Snake & $\cdots$ & Public catalogue identifier (0--1179)\tabularnewline
2 & Nstar & $\cdots$ & Number of catalogue members, including bridge stars\tabularnewline
3 & Nnode & $\cdots$ & Number of retained base nodes\tabularnewline
4 & GLON & degree & Median member Galactic longitude\tabularnewline
5 & GLAT & degree & Median member Galactic latitude\tabularnewline
6 & X & pc & Median heliocentric member $X$\tabularnewline
7 & Y & pc & Median heliocentric member $Y$\tabularnewline
8 & Z & pc & Median heliocentric member $Z$\tabularnewline
9 & logt & dex & Member-count-weighted median base-node $\log t$$^a$\tabularnewline
10 & logt\_span & dex & Weighted 84th--16th percentile base-node age range$^a$\tabularnewline
11 & FeH & dex & Member-count-weighted median base-node $[\mathrm{Fe/H}]$$^a$\tabularnewline
12 & FeH\_span & dex & Weighted 84th--16th percentile base-node metallicity range$^a$\tabularnewline
13 & SRI & $\cdots$ & Continuous full-system SRI$^b$\tabularnewline
14 & SRI\_core & $\cdots$ & SRI of the selected KPD core, when available$^c$\tabularnewline
15 & grade & $\cdots$ & Recommended full-SRI grade: Gold / Silver / Bronze$^b$\tabularnewline
16 & KPD & $\cdots$ & 1: subdivision flagged; 0: assessed, no split; blank: unassessed$^c$\tabularnewline
17 & provenance & $\cdots$ & Catalogue assembly route$^d$\tabularnewline
18 & conflict\_reviewed & $\cdots$ & Whether conflict review was applied$^d$\tabularnewline
19 & manual\_modified & $\cdots$ & Whether the final assignment was modified$^d$\tabularnewline
20 & resolution\_action & $\cdots$ & Recorded resolution action$^d$\tabularnewline
21 & SRI\_group & $\cdots$ & Full-system SRI evidence group: A, B, or C$^b$\tabularnewline
22 & core\_group & $\cdots$ & Core SRI evidence group: A, B, or C; blank if unavailable$^c$\tabularnewline
23 & Nentity & $\cdots$ & Number of full-SRI scoring entities$^b$\tabularnewline
24 & M\_RV & $\cdots$ & Number of RV-qualified scoring entities$^b$\tabularnewline
25 & S\_mem & $\cdots$ & Spatial member-cloud continuity score$^b$\tabularnewline
26 & S\_cen & $\cdots$ & Spatial centre-discontinuity score$^b$\tabularnewline
27 & V\_mem & $\cdots$ & Tangential-velocity member-cloud continuity score$^b$\tabularnewline
28 & V\_cen & $\cdots$ & Three-dimensional velocity centre-discontinuity score$^b$\tabularnewline
29 & C\_cross & $\cdots$ & Spatial--velocity relation score$^b$\tabularnewline
30 & RUWE\_overlap & $\cdots$ & Best-match member overlap with the RUWE-selected catalogue$^e$\tabularnewline
\bottomrule
\end{tabular}

\footnotesize
\begin{flushleft}
\textit{Notes.}
$^a$ Summaries use base-node member counts as weights. The spans
describe between-node variation, not uncertainty in the median.\newline
$^b$ SRI is a continuous structural-compatibility score.
Gold, Silver, and Bronze are recommended empirical grades with
boundaries at 0.70 and 0.50, not calibrated association probabilities.
Counts and component scores refer to the full structure;
A/B/C is determined by the RV-qualified entity count.
Inactive components are blank: \(S_{\rm cen}\) and \(V_{\rm cen}\)
in Group A, and \(C_{\rm cross}\) in Groups A and B.\newline
$^c$ KPD and core scores do not alter membership or the full-SRI
grade. Core fields are blank unless a unique, scorable core is
available. The recommended primary sample is Gold with assessed
\({\rm KPD}=0\); Silver may extend it after targeted checks.
Flagged systems require partition-level inspection regardless
of grade; unassessed systems are distinct from \({\rm KPD}=0\).\newline
$^d$ The provenance fields record the catalogue-assembly path and
subsequent conflict handling.\newline
$^e$ \texttt{RUWE\_overlap} is
\(\max_j |M_i\cap R_j|/\min(|M_i|,|R_j|)\),
where \(M_i\) and \(R_j\) are the member-source sets of the
adopted Snake and a RUWE-selected comparison structure,
including bridge stars. Zero denotes no shared members;
unity denotes complete containment of the smaller set,
allowing different structure boundaries.
\end{flushleft}
\end{table*}

\begin{table*}[htbp]
\nolinenumbers
\centering
\caption{Columns of the Stellar Snake Node Parameter Catalogue}
\label{table:snake_node}
\begin{tabular}{c l l l}
\toprule\toprule
 & Column & Unit & Description\tabularnewline
\midrule
1  & Snake       & $\cdots$        & Identifier of the parent Snake complex\tabularnewline
2  & id\_node    & $\cdots$        & Identifier of the base node\tabularnewline
3  & N           & $\cdots$        & Number of member stars in the node\tabularnewline
4  & sigma       & $\cdots$        & Mahalanobis overdensity significance $\sigma_{\rm obs}$\tabularnewline
5  & X           & pc              & $X$-coordinate of the node center\tabularnewline
6  & Y           & pc              & $Y$-coordinate of the node center\tabularnewline
7  & Z           & pc              & $Z$-coordinate of the node center\tabularnewline
8  & Va          & km s$^{-1}$     & Node-median tangential velocity $V_{\alpha*}$\tabularnewline
9  & Vd          & km s$^{-1}$     & Node-median tangential velocity $V_{\delta}$\tabularnewline
10 & RV          & km s$^{-1}$     & Node-level radial velocity\tabularnewline
11 & e\_RV       & km s$^{-1}$     & Uncertainty in radial velocity\tabularnewline
12 & E           & km$^2$ s$^{-2}$ & Orbital energy $E$$^a$\tabularnewline
13 & LZ          & kpc km s$^{-1}$ & Vertical angular momentum $L_Z$$^a$\tabularnewline
14 & logt        & dex             & Logarithmic age $\log t$$^b$\tabularnewline
15 & e\_logt     & dex             & Uncertainty in $\log t$\tabularnewline
16 & FeH         & dex             & Metallicity $[{\rm Fe/H}]$$^b$\tabularnewline
17 & e\_FeH      & dex             & Uncertainty in $[{\rm Fe/H}]$\tabularnewline
18 & AV          & mag             & Visual extinction $A_V$$^b$\tabularnewline
19 & e\_AV       & mag             & Uncertainty in $A_V$\tabularnewline
20 & DM          & mag             & Distance modulus$^b$\tabularnewline
21 & e\_DM       & mag             & Uncertainty in distance modulus\tabularnewline
22 & f\_broad    & $\cdots$        & Photometric broadening factor $f_{\rm broad}$$^b$\tabularnewline
23 & e\_f\_broad & $\cdots$        & Uncertainty in $f_{\rm broad}$\tabularnewline
24 & f\_cl       & $\cdots$        & PointNet-inferred photometric purity proxy for \(G<18\)~mag stars$^b$\tabularnewline
25 & e\_f\_cl    & $\cdots$        & Uncertainty in $f_{\rm cl}$\tabularnewline
\bottomrule
\end{tabular}
\footnotesize
\begin{flushleft}
\textit{Note:} $^a$ Computed in the solar-frame and potential convention of
Section~\ref{sec:data} ($R_0=8.27$~kpc, $V_c=238$~km~s$^{-1}$,
MWPotential2014-like).\\
$^b$ Inferred by the PointNet regressor (Section~\ref{sec:pointnet}).
The six corresponding \(e_*\) columns report internal Monte Carlo
standard deviations from member resampling and dropout, not externally
calibrated total uncertainties. These PointNet errors and the
node-level \(e_{\rm RV}\) are not substitutes for the source-level
astrometric and RV uncertainties required by the SRI. Entity velocities
and their error model are recomputed from Table~\ref{table:snake_member}.
\end{flushleft}
\end{table*}

\begin{table*}[htbp]
\nolinenumbers
\centering
\caption{Columns of the Stellar Snake Member-Star Catalogue}
\label{table:snake_member}
\small
\begin{tabular}{c l l l}
\toprule\toprule
 & Column & Unit & Description\tabularnewline
\midrule
1 & Snake & $\cdots$ & Public identifier of the parent Snake\tabularnewline
2 & id\_part & $\cdots$ & Identifier of the retained FoF part; present for node members and bridge stars$^c$\tabularnewline
3 & id\_node & $\cdots$ & Identifier of the retained base node; blank only for bridge stars$^c$\tabularnewline
4 & source\_id & $\cdots$ & Gaia DR3 source identifier$^a$\tabularnewline
5 & ra & degree & ICRS right ascension$^a$\tabularnewline
6 & dec & degree & ICRS declination$^a$\tabularnewline
7 & parallax & mas & Gaia DR3 parallax$^a$\tabularnewline
8 & parallax\_error & mas & Standard uncertainty of the parallax$^a$\tabularnewline
9 & pmra & mas yr$^{-1}$ & Proper motion in right ascension, $\mu_{\alpha *}=\dot{\alpha}\cos\delta$$^a$\tabularnewline
10 & pmdec & mas yr$^{-1}$ & Proper motion in declination, $\mu_\delta$$^a$\tabularnewline
11 & pmra\_error & mas yr$^{-1}$ & Standard uncertainty of $\mu_{\alpha *}$$^a$\tabularnewline
12 & pmdec\_error & mas yr$^{-1}$ & Standard uncertainty of $\mu_\delta$$^a$\tabularnewline
13 & parallax\_pmra\_corr & $\cdots$ & Correlation coefficient between parallax and $\mu_{\alpha *}$$^a$\tabularnewline
14 & parallax\_pmdec\_corr & $\cdots$ & Correlation coefficient between parallax and $\mu_\delta$$^a$\tabularnewline
15 & phot\_g\_mean\_mag & mag & Gaia DR3 mean $G$-band magnitude$^a$\tabularnewline
16 & bp\_rp & mag & Gaia DR3 colour $G_{\rm BP}-G_{\rm RP}$$^a$\tabularnewline
17 & radial\_velocity & km s$^{-1}$ & Gaia DR3 radial velocity; blank if unavailable$^a$\tabularnewline
18 & radial\_velocity\_error & km s$^{-1}$ & Standard uncertainty of the radial velocity; blank if unavailable$^a$\tabularnewline
19 & X & pc & Heliocentric Cartesian coordinate toward the Galactic centre$^b$\tabularnewline
20 & Y & pc & Heliocentric Cartesian coordinate in the direction of Galactic rotation$^b$\tabularnewline
21 & Z & pc & Heliocentric Cartesian coordinate toward the North Galactic Pole$^b$\tabularnewline
22 & E & km$^2$ s$^{-2}$ & Orbital energy in the adopted Galactic potential, when available$^b$\tabularnewline
23 & LZ & kpc km s$^{-1}$ & Vertical component of angular momentum, when available$^b$\tabularnewline
24 &ruwe & $\cdots$ & Gaia DR3 renormalised unit weight error$^a$\tabularnewline
\bottomrule
\end{tabular}

\footnotesize
\begin{flushleft}
\textit{Notes.}
$^a$ Gaia DR3 astrometric, photometric, and spectroscopic quantities
\citep{vallenari2023gaia}. \texttt{source\_id} is retained as the
original integer identifier. No RUWE cut is applied to the released
member sample; 43,133 of its 692,408 sources (6.23\%) have
\(\mathrm{RUWE}>1.4\). RUWE is provided for user-defined
astrometric-quality selections.\newline
$^b$ \(X,Y,Z\) are heliocentric Cartesian coordinates.
The orbital quantities \(E\) and \(L_Z\) are calculated using the
solar-frame and Galactic-potential conventions adopted in
Section~\ref{sec:data}.\newline
$^c$ Bridge stars retain valid \texttt{Snake} and \texttt{id\_part}
identifiers but have a blank \texttt{id\_node}.
\end{flushleft}
\end{table*}

\end{document}